\documentclass[twocolumn]{aastex631}
\usepackage{amsmath,amssymb, amsthm,amstext}
\usepackage{natbib}
\usepackage{graphicx}
\usepackage{color}
\usepackage{array, enumerate}
\usepackage{bm}
\usepackage{multirow}
\usepackage{braket}
\usepackage{txfonts}
\usepackage{physics}
\usepackage[normalem]{ulem}

\def\be{\begin{equation}}
\def\ee{\end{equation}}

\def\apjl{Astroph.J.Lett.}
\def\apjs{Astroph.J.S.}

\def\aap{Astronomy\&Astrophysics}

\begin{document}

\title{Secular evolution of quasi-periodic eruptions  }

\author[0009-0004-4654-1328]{Cong Zhou}
\email{dysania@mail.ustc.edu.cn}
\affiliation{CAS Key Laboratory for Research in Galaxies and Cosmology, Department of Astronomy, University of Science and Technology of China, Hefei 230026, People’s Republic of China}
\affiliation{School of Astronomy and Space Sciences, University of Science and Technology of China, Hefei 230026, People’s Republic of China}
\author{Yuhe Zeng}
\affiliation{Tsung-Dao Lee Institute, Shanghai Jiao-Tong University, Shanghai, 520 Shengrong Road, 201210, People’s Republic of China}
\affiliation{School of Physics \& Astronomy, Shanghai Jiao-Tong University, Shanghai, 800 Dongchuan Road, 200240, People’s Republic of China}
\author[0000-0001-9608-009X]{Zhen Pan}
\email{zhpan@sjtu.edu.cn}
\affiliation{Tsung-Dao Lee Institute, Shanghai Jiao-Tong University, Shanghai, 520 Shengrong Road, 201210, People’s Republic of China}
\affiliation{School of Physics \& Astronomy, Shanghai Jiao-Tong University, Shanghai, 800 Dongchuan Road, 200240, People’s Republic of China}

\begin{abstract}
 Quasi-periodic eruptions (QPEs) are intense repeating soft X-ray bursts with recurrence times about a few hours
to a few weeks from  galactic nuclei.  More and more analyses show that QPEs are the result of collisions 
between a stellar mass object (SMO, a stellar mass black hole or a main sequence star) and  
an accretion disk around a supermassive black hole (SMBH) in galactic nuclei.  
In this work, we propose an osculating trajectory method for efficiently calculating secular evolution of extreme mass ratio inspirals (EMRIs)
that are perturbed by an accretion disk. This method accelerates the calculation of EMRI orbital evolution by orders of magnitude and lays the foundation for analyzing long-term QPE observations. Applying this method to  orbital analyses of  GSN 069 and eRO-QPE2, the two most stable QPE sources,  
we find 
informative constraints on the SMBH mass, the radiation efficiency of QPEs, the SMO nature, the accretion disk surface density and the accretion disk viscosity.
Combining all the QPE sources available, we find the QPE EMRIs can be divided into two populations according to their orbital eccentricities,
where the orbital periods and the SMBH masses in the low-eccentricity population
follow a scaling relation $T_{\rm obt}\propto M_{\bullet}^n$ with $n\approx 0.8$. 
\end{abstract}


\section{Introduction}

Quasi-periodic eruptions (QPEs) are intense repeating soft X-ray bursts with recurrence times about a few hours
to a few weeks from  galactic nuclei. 
Since the first detection  more than a decade ago~\citep{Sun2013}, QPEs  from about ten different nearby galactic nuclei 
have been reported~\citep{Miniutti2019,Giustini2020,Arcodia2021,Arcodia2022,Chakraborty2021,Evans2023,Guolo2024,Arcodia2024,nicholl2024}.
QPEs are similar in their electromagnetic emissions,
with the peak luminosity ($10^{42}-10^{43}$ ergs s$^{-1}$),
the thermal-like X-ray spectra with temperature $kT \simeq 100-250$ eV, 
the temperature  $50-80$ eV in the quiescent state, and mild asymmetries between
the fast rise and slower decay phases in the flare light curves.

There are several pieces of observational evidence suggesting a strong connection between QPEs and tidal disruption events (TDEs). Both QPEs and TDEs are preferentially found in low-mass poststarburst galaxies
which harbor a low-mass 
($\simeq 10^5-10^7 M_\odot$) central supermassive black holes (SMBH)~\citep{Shu:2017zjd,Wevers2022,Miniutti2023},
and an extended emission line region (EELR)~\citep{Wevers:2024yws,Wevers:2024hzp}. 
Similar to TDE host galaxies, QPE hosts are of similar morphological properties that are rarely found in 
broader galaxy population~\citep{Gilbert:2024ywq}. 
A stronger QPE-TDE connection is established from the  association of  QPE sources (GSN 069,  XMMSL1 J024916.6-04124, eRO-QPE3, AT2019qiz
and a candidate AT 2019vcb)  with
previous TDEs~\citep{Shu2018,Sheng2021, Miniutti2023,Quintin2023,Arcodia2024,nicholl2024,Bykov:2024ogf}.
In particular, QPEs have been directly detected in X-ray light curves of two TDEs $\mathcal{O}(1)$ years after their ignitions~\citep{nicholl2024,Bykov:2024ogf}. 

There is also observational evidence suggesting a likely connection between QPEs/TDEs and a recent accretion phase of the central SMBH.
The presence of a narrow line region and the absence of luminous broad emission lines in
most QPE host galaxies implies the hosts are recently switched-off active galactic nuclei (AGNs),
and a long-lived accretion flow likely plays a role in the QPE phenomenon~\citep{Wevers2022}.
The high occurrence of EELRs in both TDE hosts and QPE hosts further implies high (detection) rates of TDEs and QPEs 
in  recently switched-off AGNs~\citep{French2023,Wevers:2024hzp,Wevers:2024yws}.

In addition to these common properties of QPE emission spectra and QPE host galaxies, QPE light curves also share some 
intriguing features that may reveal the QPE origin(s), including the alternating peak luminoisities $I_{\rm strong}$ and $I_{\rm weak}$
and the alternating recurrence times $T_{\rm long}$ and $T_{\rm short}$. 
For example, in the most famous QPE source GSN 069,  both $T_{\rm long}$ and $T_{\rm short}$ show large variations, while
$T_{\rm long}+T_{\rm short}$ is approximately a constant~\citep{Zhou2024a,Zhou2024b} (paper I and II hereafter). This observation clearly shows that the period of the underlying dynamical process  is  actually $T_{\rm long}+T_{\rm short}$ rather than $T_{\rm long}$ or $T_{\rm short}$, and two flares with non-uniform intervals are produced in one dynamical period.
Besides these common properties shared by most QPEs, a number of peculiar features in several QPE sources have been revealed by long term observations,
including the disappearance and reappearance of QPEs and their association with the quiescent state luminosity, the large change in the  QPE recurrence times $T_{\rm long, short}$,
the complex rising and decay profiles of QPE light curves and the non-uniform decay in the flare recurrence times (see \citealt{Miniutti2023b,Miniutti:2024rlj} for GSN 069,
\citealt{Arcodia2022,Pasham2024,Chakraborty2024} for eRO-QPE 1, \citealt{Arcodia:2024taw,Pasham:2024sox} for eRO-QPE 2
and \citealt{Giustini:2024dyy} for RX J1301).

These observations provide informative clues on the origin(s) of QPEs. A natural explanation is that 
QPEs are the result of collisions between a stellar mass object (SMO, a stellar mass black hole or a main sequence star) orbiting
around a SMBH in galactic nuclei and  an accretion disk, which may be fed by a tidal disruption event or in some cases a recently turn-on AGN disk (see the analysis of  Swift J023017 in paper II). 
Though this extreme mass ration inspiral(EMRI)+disk model is not the only interpretation proposed,  it is favored by 
more and more analyses (see e.g., \citealt{Xian2021,Franchini2023,Tagawa:2023fpb,Arcodia2024,Guolo2024,Zhou2024a,Zhou2024b,
Chakraborty2024,Linial:2024mdz,Arcodia:2024taw,Yao:2024rtl,Giustini:2024dyy,Vurm:2024vfb,Pasham:2024sox} for details). 
If the EMRI+disk interpretation is correct, QPEs will be  a sensitive probe to orbits of SMOs in the vicinity of SMBHs,
the SMBH masses,
and consequently to formation processes of EMRIs~\citep{Zhou2024a, Zhou2024b}
and
their formation rates~\citep{Arcodia:2024efe,Kaur:2024ofj}. 

In previous analyses of the orbits of QPE EMRIs, 
we considered the simplest scenario in which the SMO moves along a geodesic and the accretion disk lies on the equator~\citep{Zhou2024a, Zhou2024b}. 
We also noticed that these assumptions might be violated in the long run: e.g., we found the posterior of the EMRI orbital period 
$T_{\rm obt}$ in GSN 069 is multi-peaked, which indicates a slow but measurable evolution in the orbital period; 
we found a clear increase in the orbital period of eRO-QPE1 between Aug.
2020 and Aug. 2021, and we supposed the origin of the apparent orbital period increase is due to nodal precession and alignment of the accretion disk;
we also found a clear decrease in the orbital period of eRO-QPE2 between Aug. 2020 and Jun. 2022 and we attributed this apparent orbital period 
decrease as the SMO orbital energy dissipation as crossing the disk. Recent observations show that  apparent orbital period 
decrease of eRO-QPE2 is non-uniform, thus the modulation of the disk precession seems to play a role~\citep{Arcodia:2024taw}.

In this work, we extend the previous analyses by including extra effects, 
the SMO orbital decay due to collisions with the disk, the disk precession and possible alignment of an initially misaligned disk.
We find clear Bayes evidence for orbital decay in GSN 069 and evidence for orbital decay or disk alignment in eRO-QPE2, the two most stable  QPE sources. 
Taking these extra effects into account, we can fit multiple observations spanning  a long time with the QPE timing model.
As a result, we obtain tighter constraints on the SMO orbital parameters than in previous analyses, including the orbital size, the orbital eccentricity and the central SMBH mass.
Consistent with previous analyses,
we find the two QPE EMRIs are of nearly circular orbits with  eccentricity $e=\mathcal{O}(10^{-2})$ 
and semi-major axis $A=\mathcal{O}(10^2) M_\bullet$, where $M_\bullet$ the gravitational radius of the central SMBH.
These orbital parameters are consistent with the wet EMRI formation channel prediction~\citep{Sigl2007,Levin2007,Pan2021prd,Pan2021b,Pan2021,Pan2022,Derdzinski2023,Wang2023,Wang2023b}, 
but are incompatible with either the dry channel~\citep{Hopman2005,Preto2010,Bar-Or2016,Babak2017,Amaro2018,Broggi2022} or the Hills channel~\citep{Miller2005,Raveh2021}.
Collecting 8 QPE sources in total with reasonable measurements of the orbital period $T_{\rm obt}$ and the SMBH mass $M_\bullet$,
we find  a likely correlation relation $T_{\rm obt}\propto M_\bullet^{0.8}$ among 6 QPE sources wherein the EMRIs are of low eccentricities.

This paper is organized as follows. In Section~\ref{sec:model}, we introduce the EMRI+disk model.
In Section~\ref{sec:analyses}, we show the detailed analyses of the two QPE sources.
This paper is concluded with Section~\ref{sec:conclusions}.
Throughout this paper, we use the geometrical units with convention $G=c=1$.

\section{EMRI+Disk model}\label{sec:model}

In an EMRI+disk system, one can in principle predict the SMO-disk collision time and the resulting QPE light curve.
But the  prediction of the light curve is subject to large uncertainties in the disk model, the nature of the SMO and the radiation mechanism.
Following papers I and II,  we choose to constrain the EMRI kinematics and 
the QPE emission separately for mitigating the impact of these uncertainties: 
we first fit each QPE with a simple light curve
model and obtain the starting time of each flare $t_0\pm \sigma(t_0)$ (see paper II for details of light curve fitting),
which is identified as the observed disk crossing time and used
for constraining the EMRI orbital parameters assuming a flare timing model.

The flare timing model consists of two major components: the SMO motion and the disk motion.
In papers I and II, we have assumed that the SMO moves along a geodesic ignoring the small orbital energy dissipation due to collisions with the disk.
As noticed in paper II, the geodesic assumption may be violated in some QPE sources in the long run, where long-term observations are available
and the small orbital period decay $\dot T_{\rm obt}$ may be detectable. In this work, we consider both EMRI geodesics and forced EMRI 
trajectories where the orbital decay caused by the SMO-disk collisions is taken into account. In papers I and II, we have assumed that the disk lies on the equator, 
therefore no nontrivial disk motion. Again in paper II, we found an apparent orbital period increase in eRO-QPE 1, 
which is likely the result of disk precession and alignment. 
In paper II, we also analyzed another famous QPE source, eRO-QPE 2,  
and we attributed the apparent orbital period decay from XMM 1 (2020-08-06) to XMM 2 (2022-02-06) and 3 (2022-06-21) the SMO-disk collisions.
However, the most recent observation XMM 4 (2023-12-08) shows that the  the apparent orbital period decay is nonuniform,
and the disk precession likely plays a role here, either modulating the apparent QPE intervals or modulating the SMO orbital decay rate~\citep{Arcodia:2024taw}.
In this work, we take the precession and alignment of an initially misaligned disk into consideration.

With data $d=\{t_0^{(k)}\pm \sigma^{(k)}(t_0)\}$ ($k$ is the flare index) and a flare timing model,
we can constrain model parameters $\mathbf{\Theta}$. According to 
the Bayes theorem, the posterior of parameters is 
\be \mathcal P(\mathbf{\Theta}, \mathcal{H}|d) =\frac{ \mathcal{L}(d|\mathbf{\Theta},\mathcal{H}) \pi(\mathbf{\Theta},\mathcal{H})}{\mathcal{Z}(d)}\ , \ee
where $\mathcal{L}(d|\mathbf{\Theta},\mathcal{H})$ is the likelihood of detecting data $d$ under hypothesis $\mathcal{H}$
with model parameters $\mathbf{\Theta}$, $\pi(\mathbf{\Theta},\mathcal{H})$ is the parameter prior assumed,
and the normalization factor $\mathcal{Z}(d)$ is the evidence of hypothesis $\mathcal{H}$ with data $d$.
To quantify the support for one hypothesis $\mathcal{H}_1$ over another $\mathcal{H}_0$  by  data $d$,
we can define the Bayes factor 
\be 
\mathcal{B}_0^1 = \frac{ \mathcal{Z}_1(d)}{\mathcal{Z}_0(d)}\ .
\ee
The higher value of $\mathcal{B}_0^1$ stands for stronger support for hypothesis  $\mathcal{H}_1$ over $\mathcal{H}_0$.
According to Jeffreys's scale, $\log\mathcal{B}_0^1\in (1.2, 2.3), (2.3, 3.5), (3.5, 4.6), (4.6, \infty)$ are the criteria of substantial, strong, very strong,
and decisive strength of evidence, respectively.

In the following subsections, we will explain two main components of the flare timing model and define the likelihoods.

\subsection{EMRI trajectories}

If the SMO orbital energy loss is negligible, the SMO simply moves along a geodesic. 
Otherwise, the SMO moves along a forced trajectory where the extra force arising from collisions with the accretion disk should be taken
into account. In this work, we consider both cases and test them against QPE observations.

\subsubsection{EMRI geodesics}

The orbit of a test particle in the Kerr spacetime 
\be ds^2 = g_{\mu\nu} dx^\mu dx^\nu\ , \ee  
can be obtained by solving the geodesic equation of motion (EoM),
\be 
\dv[2]{x^\mu}{\tau}+ \Gamma^{\mu}_{\alpha\beta} \dv{x^\alpha}{\tau}\dv{x^\beta}{\tau} = 0\ ,
\ee 
where $\tau$ is the proper time and $\Gamma^{\mu}_{\alpha\beta}$ is the Christoffel connection 
(we will use the Boyer-Lindquist coordinates for the Kerr metric in this work).
Equivalently, one can recast the geodesics EoM above as a set of Hamiltonian EoMs as did in papers I and II.
The problem of the geodesic equation is the low efficiency in evolving the EMRI for a large number of cycles, especially 
when the environmental perturbation from the accretion disk makes a difference, e.g., 
observations of eRO-QPE2 during 2020-2024 span $\sim 10^4$ orbital periods and show clear evidence of orbital decay \citep{Arcodia:2024taw}.
Solving the geodesic equation or post-Newtonian EoMs are way too slow for full orbital analysis of long term observations of eRO-QPE2 like sources. In this work, we propose an osculating orbital method which accelerates the calculation of EMRI orbital evolution by orders of magnitude, therefore lays the foundation for analyzing long term QPE observations. 

We start from analytic solution to Kerr geodesics ~\citep{Fujita:2009bp,vandeMeent:2019cam}. 
In terms of  Mino time (${\rm d}\lambda=1/\Sigma \ {\rm d}\tau$)~\citep{mino},  
the equations of motion in  the radial and polar direction are decoupled
\be 
\begin{aligned}
    \left(\frac{{\rm d}r}{{\rm d}\lambda}\right)^2 &= V_r(r)\ , \\ 
    \left(\frac{{\rm d}z}{{\rm d}\lambda}\right)^2 &= V_z(z)\ , \\
    \frac{{\rm d}t}{{\rm d}\lambda} &= \frac{r^2+a^2}{\Delta}\left(E(r^2+a^2)-aL\right)-a^2E(1-z^2)+aL\ , \\
    \frac{{\rm d}\phi}{{\rm d}\lambda} &= \frac{a}{\Delta}\left(E(r^2+a^2)-aL\right)+\frac{L}{1-z^2}-aE\ ,
\end{aligned}
\label{eq:four_mino}
\ee
where $z=\cos\theta$ and $\Sigma=r^2+a^2z^2$, $\Delta=r(r-2)+a^2$. The two potentials $V_r(r)$ and $V_z(z)$ are~\citep{vandeMeent:2019cam} 
\be 
\begin{aligned}
    V_r(r) &= [(r^2+a^2)E-aL]^2-\Delta[r^2+(L-aE)^2+C]\ , \\
    V_z(z) &= C(1-z^2)-\left[(1-z^2)(1-E^2)a^2+L^2\right]z^2\ .
\end{aligned}
\ee 
where $E, L, C$ are the integrals of motion: energy, angular momentum and Carter constant~\citep{Carter1968}.  
In \citealt{Fujita:2009bp,vandeMeent:2019cam}, the solutions are written as analytic functions of four phases $\{q_r, q_z, q_t, q_\phi\}$ which evolve linearly with the Mino time, i.e. ,
\be 
\begin{aligned}
    r(\lambda) &= r(q_r(\lambda); E, L, C)\ , &q_r(\lambda) &= \Upsilon_r\lambda+q_{r,\rm ini}\ , \\ 
    z(\lambda) &= z(q_z(\lambda); E, L, C) \ , &q_z(\lambda) &= \Upsilon_z\lambda+q_{z,\rm ini}\ ,\\
    t(\lambda) &= t(q_{t,r,z}(\lambda); E, L, C)\ , &q_t(\lambda) &= \Upsilon_t\lambda+q_{t,\rm ini} \ , \\
    \phi(\lambda) &= \phi(q_{\phi,r,z}(\lambda); E, L, C)\ , &q_\phi(\lambda) &= \Upsilon_\phi\lambda+q_{\phi,\rm ini}\ ,\\
\end{aligned}
\ee 
(see \citealt{vandeMeent:2019cam} for the explicit expressions of the mino time frequencies $\Upsilon_\mu(E,L,C)$ and $x^\mu(q)$  \footnote{There a sign typo in equation (33) of \citealt{vandeMeent:2019cam}.}),
where the four frequencies $\{\Upsilon_r, \Upsilon_z, \Upsilon_t, \Upsilon_\phi\}$ are functions of integrals of motion and therefore are constant themselves, and $\{q_{r,\rm ini}, q_{z,\rm ini}, q_{t,\rm ini},q_{\phi,\rm ini}\}$ are the initial phases. As in paper II, we use the orbital parameters semilatus rectum $p$, eccentricity $e$ and 
minimum polar angle  $\theta_{\rm min}$ to label the bound Kerr geodesics.
The conversion relations between the integrals of motion $(E, L, C)$ and the  orbital parameters $(p, e, \theta_{\rm min})$ have been derived in \citealt{Schmidt2002}.

Considering a bound orbit with parameters $(p, e, \theta_{\rm min})$,
we first obtain the integrals of motion $(E, L, C)$ using the conversion relations~\citep{Schmidt2002}, 
then set the three initial phases $\{q_{r,\rm ini}, q_{z,\rm ini}, q_{\phi,\rm ini}\}$,
while the initial phase in the time direction $q_{t, \rm ini}$ is obtained from $t(q_{t, \rm ini}, q_{r, \rm ini}, q_{z, \rm ini}) =t_{\rm ini}$,
 where $t_{\rm ini}$ is some starting time point and we set $t_{\rm ini}$ as (a moment shortly before) the starting time of the first observed flare $t_0^{(1)}$ in our QPE analyses. 

\subsubsection{Forced EMRI trajectories}
\label{sec:forced_EMRI}

Taking the orbital energy dissipation into account as the SMO crosses the accretion disk into account,
the SMO does not move along a geodesic any more, i.e., the orbital parameters $\{p ,e, \cos\theta_{\rm min}\}$ 
are now time dependent. 
Similar to the geodesic case, the forced EMRI trajectories can also be computed making use of the analytic formula in the following way,
\be 
\begin{aligned}
    r(\lambda) &= r(q_r(\lambda); E, L, C)\ , &\dv{q_r}{\lambda} &= \Upsilon_r(E,L,C)\ , \\ 
    z(\lambda) &= z(q_z(\lambda); E, L, C) \ , &\dv{q_z}{\lambda} &= \Upsilon_z(E,L,C)\ ,\\
    t(\lambda) &= t(q_{t,r,z}(\lambda); E, L, C)\ , &\dv{q_t}{\lambda} &= \Upsilon_z(E,L,C) \ , \\
    \phi(\lambda) &= \phi(q_{\phi,r,z}(\lambda); E, L, C)\ , &\dv{q_\phi}{\lambda} &= \Upsilon_\phi(E,L,C) \ ,\\
\end{aligned}
\ee 
except that  $\{E, L , C\}$ and the frequencies $\{\Upsilon_r, \Upsilon_z,\Upsilon_\phi\}$ are now time dependent.

For the SMO-disk collision, the relative changes in orbital parameters are similar in magnitudes with $\delta e/e \sim \delta T_{\rm obt}/T_{\rm obt}\sim \delta \theta_{\rm min}/\theta_{\rm min}$ \cite[e.g.,][]{Linial2023,Wang2023}. As we will see later, 
the small fractional change in the orbital period $\delta T_{\rm obt}$ is detectable, while $\delta e$ and $\delta \theta_{\rm min}$ are undetectable
for the QPE sources available, because $T_{\rm obt}$ is the best constrained orbital parameter.
Therefore, we can safely take $\dot e = \dot \theta_{\rm min}=0$ in solving the EoMs above.
As for the orbital period decay rate $\dot T_{\rm obt}(t)$,  we consider 
\be \label{eq:vary_Tdot}
\dot T_{\rm obt}(t) = \dot T_{\rm obt, max} \sin\iota_{\rm sd}(t)\ ,
\ee
where $\iota_{\rm sd}$ the angle between the SMO orbital plane and the disk plane (see Eq.~[\ref{eq:delta_E_star}]).
The former is a leading-order approximation for any slowly varying orbital decay rate.
For a precessing misaligned disk, $\iota_{\rm sd}$ is expected to be modulated by both the disk precession and the SMO orbital precession, therefore the decay rate $\dot T_{\rm obt}$ is non-uniform.
For a disk on the equator, $\iota_{\rm sd}$ is a constant, then the orbital period decay rate is also a constant.

A similar osculating trajectory method has been widely used in modeling LISA EMRI waveform, where LISA EMRI orbital decay is from gravitational wave emission dissipation, while QPE EMRI orbital decay studied in this work is from collisions with accretion disks. Comparing to numerically integrating the 
geodesic equation where the shortest timescale is the orbital period $T_{\rm obt}$, the shortest timescale in the osculating trajectory method is the 
much longer orbital decay timescale $T_{\rm obt}/|\dot T_{\rm obt}| \gg T_{\rm obt}$. This is the origin of speed-ups  in computing EMRI trajectories in this work. This method lays the foundation of analyzing long-term QPE observations.

\subsection{Disk precession and alignment}

As a minimal assumption of the disk precession, we model it as a rigid body like precession with a constant precession rate. The normal vector of the disk plane is $\vec n_{\rm disk}=(\sin\beta\cos\alpha, \sin\beta\sin\alpha, \cos\beta)$, where $\alpha\in (0, 2\pi)$ is the azimuth angle and $\beta\in (0, \pi/2)$ is the angle between the disk plane and the equatorial plane. The azimuth angle then evolves as 
\be 
\begin{aligned}
    \alpha(t) &= \alpha_{\rm ini}+\frac{2\pi}{\tau_{\rm p}}(t-t_{\rm ini})\ , \\
\end{aligned}
\ee 
where $\tau_{\rm p}$ is the disk precession period, and $\alpha_{\rm ini}$, $\beta_{\rm ini}$ are the initial values of the azimuth and the polar angles specifying the  disk orientation at $t_{\rm ini}$. 

There is no obvious function form of parameterizing the disk alignment process, i.e., $\beta(t)$.
One can in principle constrain $\beta(t)$ in a non-parametric approach
with dense observations of the QPEs, which however are not available for the QPE sources considered in this work.
Limited by the amount of information available, we consider two extremal alignment processes: 1) slow alignment where $\beta(t)=\beta_{\rm ini}$;
2) fast alignment where $\beta = \beta_{\rm ini}$ during the 1st observation and $\beta = 0$ during subsequent observations.

We identify  the disk crossing time as when the SMO crosses the upper surface or the lower surface of the disk depending on the observer direction,
 i.e. $r_{\rm crs}(\vec n_{\rm crs}\cdot \vec n_{\rm disk})=H \ {\rm sign}(\vec n_{\rm obs}\cdot \vec n_{\rm disk})$. 
Without loss of generality, we fix the observer in the $x-z$ plane, i.e., the unit direction vector pointing to the observer is $\vec n_{\rm obs} = (\sin\theta_{\rm obs}, 0, \cos\theta_{\rm obs})$.
We can fix the observer in the upper semisphere when considering a disk lying on the equator, i.e.,  $\theta_{\rm obs}\in (0, \pi/2)$.
The disk height to the mid-plane is set as $H=1.5 M_\bullet$ as in paper II. 

\subsection{Flare timing model}

To summarize, we will consider the following  hypotheses that are slight different in the SMO motion,

\begin{itemize}
    \item Free EMRI hypothesis ($\mathcal{H}_{\rm e0}$): The SMO moves along a geodesic around the SMBH, which can be specified by $8$ parameters: the intrinsic orbital parameters $(p, e, \theta_{\rm min})$, the initial phases $(q_{r,{\rm ini}}, q_{z,{\rm ini}}, q_{\phi,{\rm ini}})$, the mass of the SMBH $M_\bullet$ or equivalently the orbital period $T_{\rm obt} :=2\pi (A/M_\bullet)^{3/2} M_\bullet$  (with semi-major axis $A = p/(1-e^2)$), the dimensionless spin of the SMBH $a$.
    \item Forced EMRI hypothesis ($\mathcal{H}_{\rm e1}$):  a forced EMRI orbit due to SMO-disk collisions with the orbital period decay rate 
    quantified by an extra parameter $\dot T_{\rm obt}$ or $\dot T_{\rm obt, max}$  [Eq.~(\ref{eq:vary_Tdot})] .
\end{itemize}

and in the disk motion,

\begin{itemize}
    \item Equatorial disk hypothesis ($\mathcal{H}_{\rm d0}$): a disk simply lies on the equator.
    
    \item Misaligned disk hypothesis ($\mathcal{H}_{\rm df/ds}$): we consider a misaligned disk with an
    initial orientation in the $(\alpha_{\rm ini}, \beta_{\rm ini})$ direction. The disk precesses with a  period $\tau_{\rm p}$
    and completes the alignment in a fast/slow way ($\mathcal{H}_{\rm df/ds}$) as explained in the previous subsection.
\end{itemize}

A full flare timing model consists of an EMRI motion hypothesis and a disk motion hypothesis, e.g., we will consider a vanilla hypothesis 
($\mathcal{H}_{\rm 0} = \mathcal{H}_{\rm e0}+\mathcal{H}_{\rm d0}$) as a reference for every QPE source analyzed in this work.

For a given SMO trajectory and the disk motion, one can calculate the disk crossing times $t_{\rm crs}$,
which we identify as the flare starting times. 
The propagation times of different flares at different collision locations $r_{\rm crs} \vec n_{\rm crs}$
to the observer will also be different. Taking the light propagation delays into account,
we can write $t_{\rm obs} = t_{\rm crs} + \delta t_{\rm geom} + \delta t_{\rm shap}$, where 
\be\label{eq:tobs}
\begin{aligned}
    \delta t_{\rm geom} &= -r_{\rm crs} \vec n_{\rm obs}\cdot \vec n_{\rm crs}\ , \\ 
    \delta t_{\rm shap} &=-2M_\bullet\log \left[r_{\rm crs} (1+ \vec n_{\rm obs}\cdot \vec n_{\rm crs})\right]\ ,
\end{aligned}
\ee 
are corrections caused by different path lengths and different Shapiro delays~\citep{Shapiro1964}, respectively.

\subsection{Flare timing likelihood}

Following paper II, we introduce a parameter $\sigma_{\rm sys}$ 
quantifying the effect of any physical processes that are relevant but not included in our flare timing model,
assuming the unmodeled advances or delays in the flare timing follows a Gaussian distribution with variance $\sigma_{\rm sys}^2$.
Similar inference method has been used in the context of hierarchical test of General Relativity with gravitational waves~\citep{Isi2019}.
As a result, the likelihood of seeing data $d=\{t_0^{(k)}\}$ under hypothesis $\mathcal{H}$ with model parameters is  
\be \label{eq:likeli}
\mathcal{L}_{\rm timing}(d|{\mathbf{\Theta}}, \mathcal{H})=\prod_{k}\frac{1}{\sqrt{2\pi (\tilde\sigma(t_0^{(k)}))^2}}
 \exp\left\{-\frac{(t_{\rm obs}^{(k)}- t_0^{(k)})^2}{2(\tilde\sigma(t_0^{(k)}))^2} \right\} \ ,
\ee 
where $(\tilde\sigma(t_0^{(k)}))^2=(\sigma(t_0^{(k)}))^2+\sigma_{\rm sys}^2$ is the uncertainty contributed by both modeled and unmodeled uncertainties. We also incorporate the central SMBH mass measurement from the stellar velocity dispersion using the $M_\bullet-\sigma_\star$ relation~\citep{Tremaine2002,Gultekin2009} as a contribution to the total likelihood as
\be 
\mathcal{L}_{M_\bullet}(d|{\mathbf{\Theta}}, \mathcal{H}) = \frac{1}{\sqrt{2\pi \sigma_{\log_{10} M_\bullet}^2}} \exp\left\{-\frac{(\log_{10} M_\bullet - \mu_{\log_{10} M_\bullet})^2}{2 \sigma_{\log_{10} M_\bullet}^2} \right\}\ ,
\ee 
where $\mu_{\log_{10} M_\bullet}$ and $\sigma_{\log_{10} M_\bullet}$ are the central value and the uncertainty of inferred SMBH mass, respectively (see \citealt{Wevers2022} for a brief summary of the mass measurements of SMBHs in the QPE host galaxies). The the total likelihood is therefore $\mathcal{L}_{\rm timing}\times\mathcal{L}_{M_\bullet}$. 
In this work,  we use the \texttt{nessai}~\citep{nessai} algorithm within \texttt{Bilby}~\citep{Ashton2019} for model parameter inference on GSN 069.  For eRO-QPE2, we employ the \texttt{pymultinest}~\citep{pymultinest} algorithm, also within \texttt{Bilby}, leveraging its MPI support.

Due to the efficiency of the osculating trajectory method explained in the previous subsection, the model parameter inference is accelerated by orders of magnitude. Taking GSN 069 as an example,  the sampling time for $\mathcal{H}_0$ is 43 minutes with 1000 live points on 64 cores in this work,
comparing to a sampling time of 3 hours and 43 minutes with 200 live points on 128 cores in paper II where we directly integrated the geodesic equation. 
This is amount to a speed-up by a factor of $52$.  For reference, it takes about 5 hours in analyzing GSN 069  ($\mathcal{H}_1$) with 1000 live points on 64 cores.
The eRO-QPE2 observations span $\sim \mathcal{O}(10^4)$ orbits, therefore take more resources to analyze.   
The sampling times are about 30 minutes, 8 days and 20 days for $\mathcal{H}_0,\ \mathcal{H}_1$ and $\mathcal{H}_2$ respectively with 640 cores across 10 nodes using Message Passing Interface (MPI).

\section{Analyses of QPE sources}\label{sec:analyses}

\subsection{GSN 069}

In addition to the 3 XMM-Newton observations used in paper I and II, to better constrain the orbital evolution, we also take the Chandra observation (ObsID:22096) at 2019-02-14 into account in this work. The Chandra data can be obtained by the Chandra X-ray Observatory, contained in~\dataset[DOI: 10.25574/cdc.368]{https://doi.org/10.25574/cdc.368}. Following paper I and \citealt{Miniutti2023b}, we process the Chandra data with CIAO software and dmextract task. The energy band used for Chandra light curve extraction is 0.4-2keV, whereas for XMM-Newton, it is 0.2-2keV. This difference arises from severe contamination degrading the low-energy response of Chandra's ACIS detector, as discussed in \citealt{Miniutti2023b}. As has been illustrated in paper I, the light curve profile of Chandra is obviously incomplete, making it hard to locate the flare starting time $t_0$. However, given the accurate determination of flare peak time $t_p$ in Chandra and also XMM-Newton light curve data, we can statistically obtain $t_p-t_0$ from the XMM-Newton data and subsequently derive $t_0$ in Chandra data.

Combining 3 XMM-Newton observations and 1 Chandra observation in 2018 and 2019, we first constrain the orbital parameters in GSN 069 under the vanilla hypothesis $\mathcal{H}_{0}$ and we find 
\be \label{eq:GSN_cons_H0}
\begin{aligned}
    p &= 289^{+138}_{-189}\  M_\bullet \ ,\\ 
    e &= 0.04^{+0.05}_{-0.04}\ , \\
    T_{\rm obt} &= 63.70^{+0.02}_{-0.07}\ {\rm ks}\ ,\\
\end{aligned}
\ee 
at $95\%$ confidence level. The posterior corner plot of all model
parameters are shown in Fig.~\ref{fig:GSN069_H0}.
Slightly different from Paper II, we have included the Chandra observation in the orbital analysis here.
Though the data quality is not as good as XMM-Newton observations, the three data points added still tighten the parameter constraints.

\begin{figure*}
\includegraphics[scale=0.42]{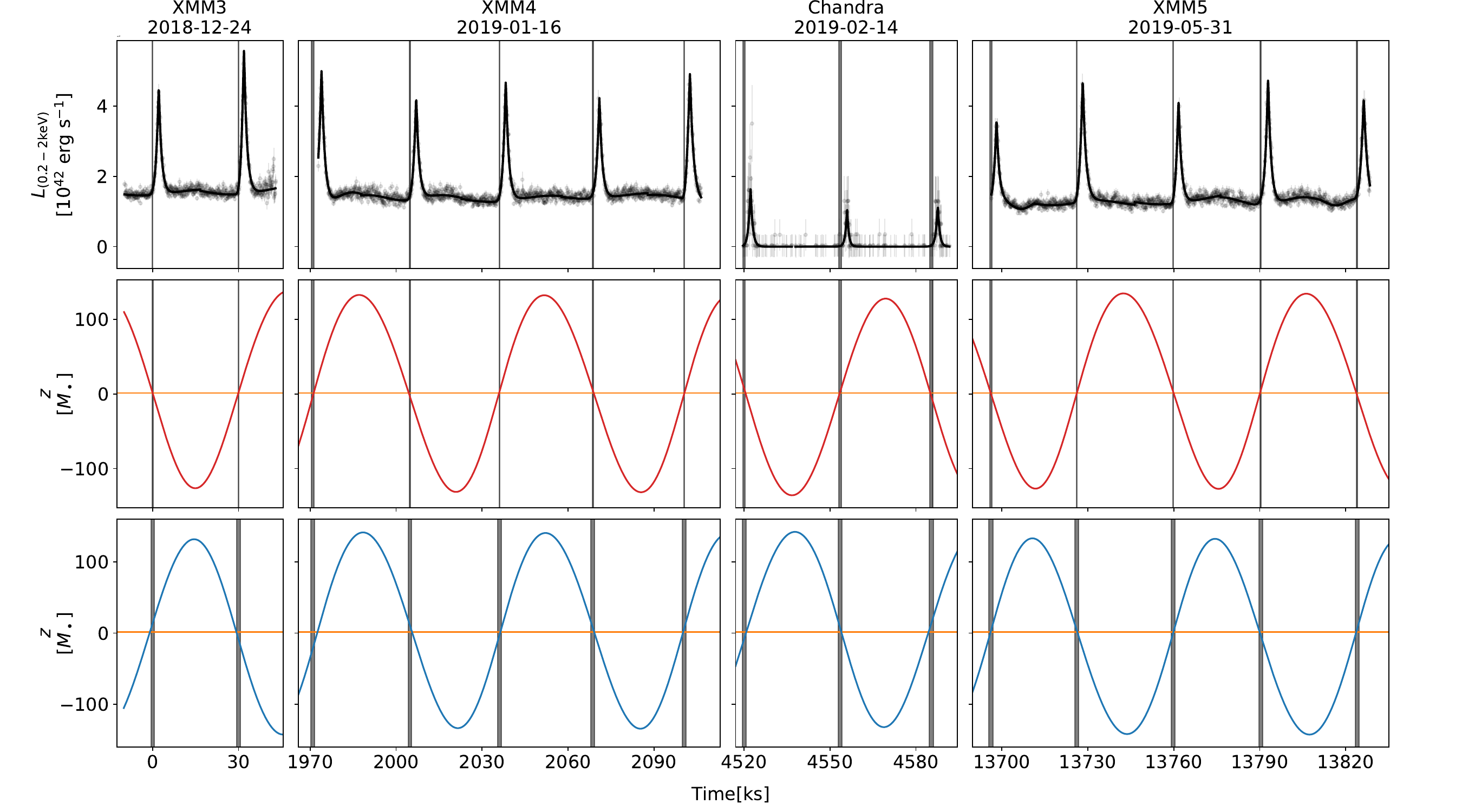}
\caption{\label{fig:lo_GSN} Top panel: light curve data of GSN 069 along with the best-fit EMRI trajectories, where the vertical bands are the inferred starting times $t_0^{(k)}\pm \sigma(t_0^{(k)})$ of the QPEs. Mid/Bottom panel: distance to the disk midplane $z_{\rm disk}(t)$ of the best-fit orbits, where the orange horizontal lines denote the disk surface $z=H$ and the verticals bands are the inferred starting times $t_0^{(k)}\pm \tilde\sigma(t_0^{(k)})$, with
$\tilde\sigma(t_0^{(k)})=\sqrt{(\sigma(t_0^{(k)}))^2+\sigma_{\rm sys}^2}$. Red: forced EMRI hypothesis ($\mathcal{H}_1$). Blue: vanilla hypothesis ($\mathcal{H}_0$).
}
\end{figure*}

A natural extension to the vanilla hypothesis is the orbital decay due to  the energy loss as the SMO crosses the disk.
Under the hypothesis of a forced EMRI+an equatorial disk ($\mathcal{H}_{1}=\mathcal{H}_{\rm e1}+\mathcal{H}_{\rm d0}$), we constrain  the orbital decay rate $\dot T_{\rm obt}$ in addition to the orbital parameters as 
\be \label{eq:GSN_cons_H1}
\begin{aligned}
    p &= 306^{+48}_{-51}\ M_{\bullet},\\ 
    e &= 0.04^{+0.02}_{-0.03},\\
    T_{\rm obt} &= 64.74^{+0.02}_{-0.03}\ \mathrm{ks},\\
    \dot T_{\rm obt} &= -6.5^{+0.2}_{-0.2}\times 10^{-5}\ ,
\end{aligned}
\ee 
at $95\%$ confidence level. 
The posterior corner plot of all model parameters are shown in Fig.~\ref{fig:GSN069_H1}. 
We also find the Bayes factor between the two hypotheses
\be 
\log \mathcal{B}_0^1= 4.4\pm 0.2\ ,
\ee 
at $95\%$ confidence level, 
which is  decisive evidence for the non-zero orbital decay. 
For comparison, the two best-fit orbits under the two hypotheses are shown in Fig.~\ref{fig:lo_GSN}, 
where the systematic uncertainty $\sigma_{\rm sys}$ in $\mathcal{H}_1$ is much lower than in $\mathcal{H}_0$.
This is another evidence for the non-zero orbital decay.
Disk precession has also been considered, 
and analysis of GSN 069 QPEs shows no signature of disk precession.

A side note is that the posterior of $p$ is actually bimodal as shown in  Fig.~\ref{fig:GSN069_H1}, with
a large peak at $p= 306^{+48}_{-51}\ M_{\bullet}$ [Eq.~(\ref{eq:GSN_cons_H1})] and a minor peak at $p= 145^{+26}_{-28} M_\bullet$ 
(at $95\%$ confidence level), respectively. The current data could not completely exclude the minor peak.

With the data favored  hypothesis $\mathcal{H}_1$,  the orbital analysis above has a number of immediate applications.

1. \emph{SMBH mass.}  One can constrain the SMBH mass from the SMO orbital period $T_{\rm obt}$ and the Schwarzschild precession period  $T_{\rm prec}$, both of which can be inferred from the QPE timing.
The orbital period  
\be\label{eq:phy1} 
T_{\rm obt}=2\pi (A/M_\bullet)^{3/2} M_\bullet 
\ee is the best constrained orbital parameter, which can be easily identified as $T_{\rm long}+T_{\rm short}$. 
The apsidal precession period $T_{\rm prec}$ is much longer with a ratio
\be\label{eq:phy2}  
\frac{T_{\rm prec}}{T_{\rm obt}} =\frac{p}{3M_\bullet}\approx \frac{A}{3M_\bullet}  \ ,
\ee 
where the approximation is accurate for low-eccentricity orbits. 
The apsidal precession period can be inferred from the modulation of $T_{\rm long}$ and $T_{\rm short}$ (see Fig.~1 in paper I).
We find that observations XMM 3-5 span  $\approx 200$ orbital periods, which
are sufficiently long to resolve the apsidal precession period, $T_{\rm prec}\approx 102 T_{\rm obt}$. Consequently, both the orbital size $p$ and the 
SMBH mass $M_\bullet$ can be constrained. As a result, we find
\be \label{eq:mass_gsn}
\mathrm{log}_{10}(M_{\bullet}/M_{\odot}) = 5.6^{+0.1}_{-0.1}\quad ({\rm high-}p)\ ,
\ee 
at $95\%$ confidence level. 

Current observations favor a long precession period $T_{\rm prec}\approx 100 T_{\rm obt} (p\approx 300 M_\bullet)$, 
but  a shorter one $T_{\rm prec}\approx 50 T_{\rm obt}  (p\approx 150 M_\bullet)$ is not completely excluded.
Taking the less favored minor peak in the posterior at $p= 145^{+26}_{-28} M_\bullet$ into account, we find 
\be 
\mathrm{log}_{10}(M_{\bullet}/M_{\odot}) = 6.1^{+0.1}_{-0.1}\quad ({\rm low-}p)\ ,
\ee 
at $95\%$ confidence level.
Both of which are consistent with and of much lower uncertainty than the constraint $\mathrm{log}_{10}(M_{\bullet}/M_{\odot}) = 6.0\pm1.0$ (at $95\%$ confidence level) inferred from the $M_\bullet-\sigma_\star$ relation~\citep{Wevers2022}. 

Note that we have included an external measurement of the SMBH mass  \citep{Wevers2022} in defining the total likelihood.
Therefore, the constraint of the SMBH mass obtained here is not entirely independent. 
The key point is the much tighter constraint after adding the QPE timing information. This shows that
the most constraining power actually is from the QPE timing instead of from the external measurement
and QPE timing as an independent probe is of great potential in precision measurement of SMBH masses.

2. \emph{QPE radiation efficiency.} 
In terms of the amount of orbital energy loss $ E_{\rm col}$ per collision, we find
\be \label{eq:E_col}
 E_{\rm col} = \frac{E_{\rm obt}}{3} \dot T_{\rm obt} =  6.3 ^{+1.3}_{-0.9}\times 10^{46}\ \left(\frac{m}{M_\odot}\right)\   \mathrm{ergs} \ ,
\ee 
where $E_{\rm obt}=-GM_\bullet m/2A$ is the SMO orbital energy.
In combination with the estimation of total energy radiated during one QPE flare~\citep{Miniutti2023b}, $E_{\rm QPE} \approx 6.7\times 10^{45}\ \mathrm{ergs}$,
we obtain a high QPE radiation efficiency 
\be \label{eq:eta}
\eta_{\rm QPE}= \frac{E_{\rm QPE}}{E_{\rm col}}\approx 11^{+2}_{-2} \% \times \left(\frac{m}{M_{\odot}}\right)^{-1}\ ,
\ee 
at $95\%$ confidence level,
which yields a tight constraint on possible QPE radiation mechanisms. 

3. \emph{Disk surface density.} Equating the measured orbital energy loss $ E_{\rm col}$ per collision to the theoretical expectation if the SMO is a normal star (paper I)
\be \label{eq:delta_E_star}
\begin{aligned}
     \delta E_{\star} 
     &= 2\times \frac{1}{2} \delta m_{\rm gas} v_{\rm rel}^2 \, \\
     &\approx  10^{46} {\rm ergs}\times \Sigma_5
     R_{\star,\odot}^2   r_{300}^{-1} \sin\iota_{\rm sd}\ ,
\end{aligned}
\ee 
we find the disk surface density 
\be \label{eq:sigma_measure}
\Sigma_5 \approx 6.3 ^{+1.3}_{-0.9} m_{\star,\odot} R_{\star,\odot}^{-2}   r_{300} (\sin\iota_{\rm sd})^{-1}\ ,
\ee 
where $\delta m_{\rm gas}$ is the amount of gas shocked by the star, $v_{\rm rel}$ is the relative velocity between the star and the local gas during collision,
$\Sigma_5=\Sigma/(10^{5}\ {\rm g\ cm^{-2}})$,
 $m_{\star,\odot} = m_\star/M_\odot$, $R_{\star,\odot} = R_\star/R_\odot$ and 
 $r_{300}=r/300 M_\bullet$.

4. \emph{Accretion disk viscosity.} 
We consider the standard thin disk model~\citep{SS1973}, 
where the disk structure in the radiation dominated regime can be analytically expressed as~\citep{Kocsis2011}
\be \label{eq:sigma_alpha}
\begin{aligned}
    \Sigma_5 
    &= 1.7 \alpha_{0.01}^{-1} \dot M_{\bullet,0.1}^{-1} r_{100}^{3/2}\ , \\ 
    &= 3.5 \left(\frac{\alpha}{0.1}\right)^{-1}\left(\frac{L_{\rm bgd}}{10^{42}\ {\rm ergs\ s^{-1}}}\right)^{-1} \left( \frac{T_{\rm obt}}{64\ {\rm ks}} \right)\ ,
\end{aligned}
\ee 
for $\alpha$-disks, where
$\dot M_{\bullet,0.1}=\dot M_\bullet/(0.1 \dot M_{\bullet,\rm Edd})$ 
with $\dot M_{\bullet,\rm Edd}$ the Eddington accretion rate, and 
we have used the background luminosity $L_{\rm bgd}=  10^{37}\ {\rm ergs\ s^{-1}} \times (M_\bullet/M_\odot) \dot M_{\bullet,0.1}$.
Combining Eqs.(\ref{eq:sigma_measure}) and (\ref{eq:sigma_alpha}), we obtain $\alpha\approx 0.05$ for the $\alpha$-disk model.

In the same way, we consider $\beta$-disks with surface density profile~\citep{Kocsis2011}
\be \label{eq:sigma_beta}
\begin{aligned}
    \Sigma_5 &=3.2 \alpha_{0.1}^{-4/5} \dot M_{\bullet,0.1}^{3/5} M_{\bullet,5}^{1/5} r_{100}^{-3/5}\ , \\
    &= 0.95 \left(\frac{\alpha}{0.1}\right)^{-4/5}\left(\frac{L_{\rm bgd}}{10^{42}\ {\rm ergs\ s^{-1}}}\right)^{3/5} \left( \frac{T_{\rm obt}}{64\ {\rm ks}} \right)^{-2/5}\ .
\end{aligned}
\ee 
Combining Eqs.(\ref{eq:sigma_measure}) and (\ref{eq:sigma_beta}), we obtain $\alpha\approx 0.01$ for the $\beta$-disk model.

5. \emph{SMO nature.} 
In the literature, a stellar mass black hole (sBH) has been also been discussed as another possibility of the SMO. 
In this case, the influence radius within which the gas in the disk is shocked is the accretion radius of the sBH rather than its geometrical size.
To produce sufficiently energetic flares, the sBH mass has to be $m\gtrsim 30 M_\odot$ and the sBH orbit has to be close to the accretion disk with 
a small inclination angle $\iota_{\rm sd} \lesssim 0.1$~\citep{Franchini2023}.
According to Eq.~(\ref{eq:eta}), the sBH interpretation favors a low radiation efficiency
\be 
\eta_{\rm QPE} \approx 0.3 \% \times \left(\frac{m}{30 M_{\odot}}\right)^{-1}\ ,
\ee 
which seems too low for either radiation from shocked gas or radiation from gas accretion by the sBH.

In the case of a sBH, the orbital energy loss per collision due to dynamical friction is estimated as (paper I)
\be \label{eq:delta_E}
\begin{aligned}
    \delta E_{\rm sBH} 
    &  
    = 4\pi \ln\Lambda  \frac{G^2m^2}{v_{\rm rel}^2}\frac{\Sigma}{\sin(\iota_{\rm sd})}\ ,\\ 
    &\approx 6\times10^{46} {\rm ergs} \left(\frac{\ln\Lambda}{10}\right) \Sigma_5 m_{30}^2 r_{300} \left(\frac{\sin\iota_{\rm sd}}{0.1}\right)^{-3} \ ,
\end{aligned}
\ee 
where $\ln\Lambda$ is the Coulomb logarithm, $m_{30} := m/30 M_\odot$. Equating $\delta E_{\rm sBH} $ to $E_{\rm col}$ inferred from the orbital 
decay rate in Eq.~(\ref{eq:E_col}),
we obtain a dense accretion disk with surface density 
\be 
 \Sigma_5  = 32^{+6}_{-4}  \left(\frac{\ln\Lambda}{10}\right)^{-1}  m_{30}^{-1} r_{300}^{-1} \left(\frac{\sin\iota_{\rm sd}}{0.1}\right)^{3}\ ,
\ee  
at $95\%$ confidence level,
which can be accommodated in an $\alpha$-disk with a low viscosity $\alpha\approx 0.01$  [Eq.~(\ref{eq:sigma_alpha})], 
but seems hard to fit in a $\beta$-disk with a reasonable viscosity coefficient [Eq.~(\ref{eq:sigma_beta})].

\begin{table}
    \centering
    \resizebox{0.8\columnwidth}{!}{%
    \begin{tabular}{l|ccc}
       $\mathbf{\Theta}$ & $\pi(\mathbf{\Theta}, \mathcal{H}_1)$ & $\pi(\mathbf{\Theta}, \mathcal{H}_0)$ &  \\
        \hline
       $p\ [M_\bullet]$ & $ \mathcal{U}[50, 500]$ &  &  \\ 
       $e$ & $\mathcal{U}[0, 0.9]$ &  & \\
       $\cos(\theta_{\rm min})$ & $\mathcal{U}[0, 1]$&  & \\
       $\chi_{r0}$ & $\mathcal{U}[0, 2\pi]$ &  & \\
       $\chi_{\theta 0}$ &$\mathcal{U}[0, 2\pi]$ &  & \\
       $\phi$ & $\mathcal{U}[0, 2\pi]$ &  & \\
       $T_{\rm obt}\ [{\rm ks}]$ & $\mathcal{U}[60, 70]$ &  & \\
       $a$ & $\mathcal{U}[0, 1]$&  & \\
       $\theta_{\rm obs}$ & $\mathcal{U}[0, \pi/2]$ & & \\
       $\sigma_{\rm sys} \ [{\rm ks}]$ & $\mathcal{U}[0, 500]$  &  & \\
       $\dot T_{\rm obt} \ [\times 10^{-5}]$ & $\mathcal{U}[-10, 0]$  & None &
    \end{tabular} }
    \caption{Priors used for the orbital parameter inference of GSN 069 EMRI. Entries left blank in the $\pi(\mathbf{\Theta}, \mathcal{H}_0)$ column indicate that the same prior values from the $\pi(\mathbf{\Theta}, \mathcal{H}_1)$ column are imposed.} 
    \label{tab:GSN_prior}
\end{table}

\subsection{eRO-QPE2}

\begin{figure*}
\includegraphics[scale=0.55]{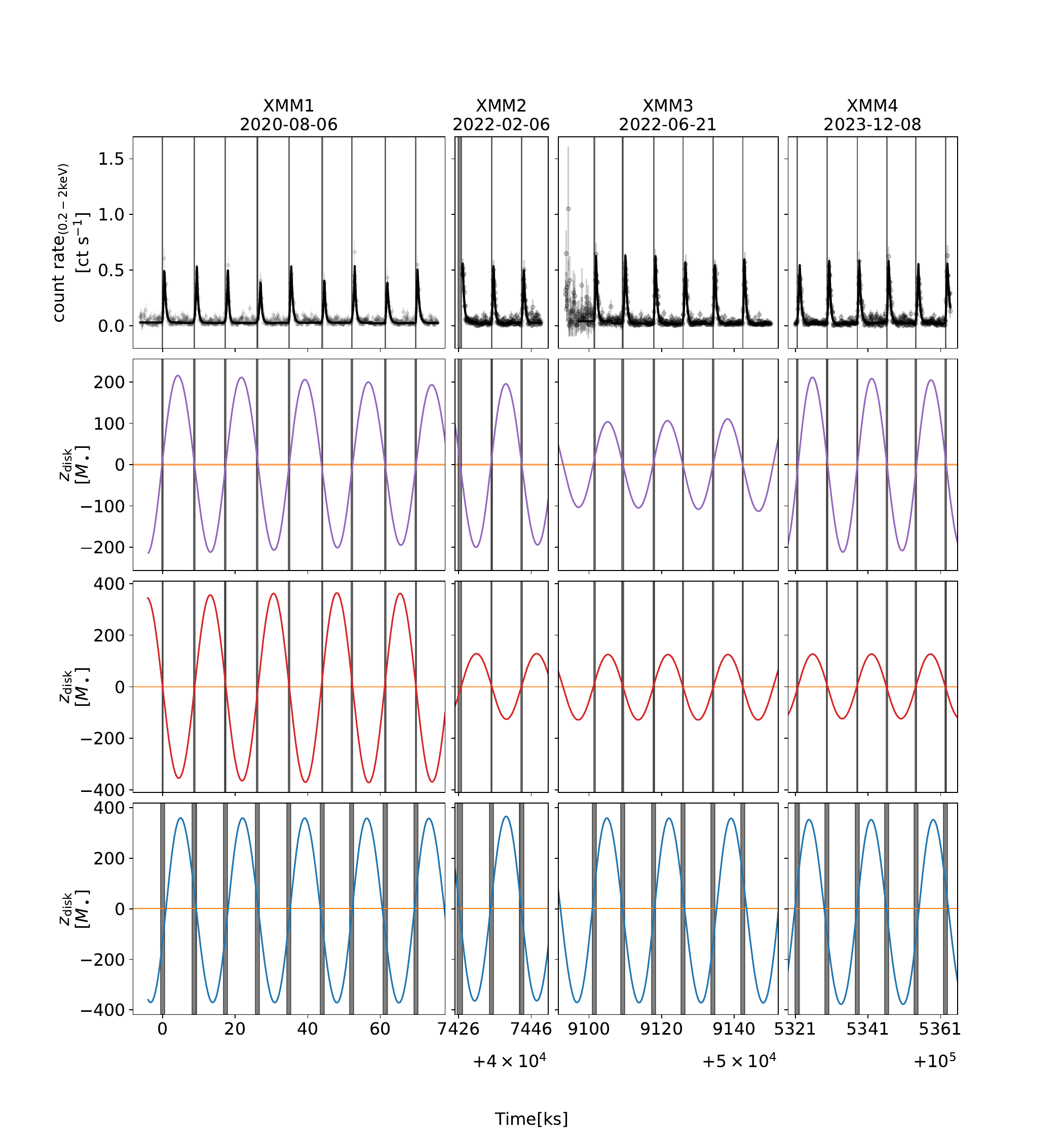}
\caption{\label{fig:lo_eRO2} Same to Fig.~\ref{fig:lo_GSN} except for eRO-QPE2. 
Purple: Forced EMRI (nonuniform) + slow alignment hypothesis ($\mathcal{H}_2$); Red: Forced EMRI (uniform) + fast alignment hypothesis ($\mathcal{H}_1$); Blue: vanilla hypothesis ($\mathcal{H}_0$). 
}
\end{figure*}

In paper II, we analyzed three public XMM-Newton observations of eRO-QPE2 at that time: XMM1 (2020-08-06), XMM2 (2022-02-06) and XMM3 (2022-06-21). We noticed a clear decreasing trend of the apparent orbital period from approximately 17.5 ks in XMM1 to 16.4 ks in XMM2 and 3. 
More observations have been done on eRO-QPE2, one of which, XMM4 (2023-12-08), is recently publicly available.
Surprisingly, there is little decay in the orbital period between XMM3 and XMM4 (Fig.~2 in \citealt{Arcodia:2024taw}).

To understand the whole evolution history of the apparent orbital period, we consider two different possibilities: 1) a hypothesis of a forced EMRI trajectory and a fast disk alignment ($\mathcal{H}_1=\mathcal{H}_{\rm e1}+\mathcal{H}_{\rm df}$), where the apparent long orbital period in XMM1 is primarily a result of a misaligned disk precessing in the same direction to the SMO orbital angular momentum direction, and the slow apparent orbital evolution in XMM2-4 is dominated by the constant $\dot T_{\rm obt}$;
2) a hypothesis of a forced EMRI trajectory and a precessing disk ($\mathcal{H}_2=\mathcal{H}_{\rm e1}+\mathcal{H}_{\rm ds}$),
where the nonuniform orbital decay is modulated by the angle between the SMO orbital plane and the disk plane
as formulated in Eq.~(\ref{eq:vary_Tdot})~\citep{Arcodia:2024taw}.
For comparison, we also consider the vanilla hypothesis  ($\mathcal{H}_0=\mathcal{H}_{\rm e0}+\mathcal{H}_{\rm d0}$) as in the analysis of GSN 069.

\begin{table}
    \centering
    \resizebox{1\columnwidth}{!}{%
    \begin{tabular}{l|ccc}
       $\mathbf{\Theta}$ & $\pi(\mathbf{\Theta},\mathcal{H}_2)$ & $\pi(\mathbf{\Theta}, \mathcal{H}_1)$ & $\pi(\mathbf{\Theta},\mathcal{H}_0)$ \\
        \hline
       $p\ [M_\bullet]$ & $ \mathcal{U}[10, 1000]$ &  &  \\ 
       $e$ & $\mathcal{U}[0, 0.9]$ &  & \\
       $\cos(\theta_{\rm min})$ & $\mathcal{U}[0, 1]$&  & \\
       $q_{r,{\rm ini}}$ & $\mathcal{U}[0, 2\pi]$ &  & \\
       $q_{z,{\rm ini}}$ &$\mathcal{U}[0, 2\pi]$ &  & \\
       $q_{\phi,{\rm ini}}$ & $\mathcal{U}[0, 2\pi]$ &  & \\
       $T_{\rm obt}\ [{\rm ks}]$ & $\mathcal{U}[14.4, 25]$ &  &  \\
       $a$ & $\mathcal{U}[0, 1]$&  & \\
       $\theta_{\rm obs}$ & $\mathcal{U}[0, \pi/2]$  &  & \\
       $\alpha_{\rm ini}$ & $\mathcal{U}[0, 2\pi]$ &  & None \\
       $\tau_{\rm p} \ [{\rm days}]$ & $\mathcal{U}[0.5, 50]$ &  & None \\
       $\beta_{\rm ini}$ & $\mathcal{U}[0, \pi/2]$ &  & None \\
       $\dot{T}_{\rm obt} \ [\times10^{-5}]$ & None & $\mathcal{U}[-10, 0]$ & None \\
       $\dot{T}_{\rm obt,max} \ [\times10^{-5}]$ & $\mathcal{U}[-10, 0]$ &  & None \\
       $\sigma_{\rm sys} \ [{\rm ks}]$ & $\mathcal{U}[0, 2]$  &  &
    \end{tabular} }
    \caption{Same to Table~\ref{tab:GSN_prior} except for eRO-QPE2.
    } 
    \label{tab:eRO2_prior}
\end{table}

In Table~\ref{tab:eRO2_prior}, we show the priors of model parameters used for orbital analyses in $\mathcal{H}_0$, $\mathcal{H}_1$ and $\mathcal{H}_2$. In Fig.~\ref{fig:lo_eRO2}, the best-fit EMRI
trajectories of the three hypotheses are displayed together with
the QPE light curves, with orbital parameters $p=572  \ M_\bullet, e=0.04, T_{\rm obt}=17.1  \ {\rm ks}$ ($\mathcal{H}_0$),  and 
 $ p= 446 \ M_\bullet, e= 0.02, T_{\rm obt}=16.9 \ {\rm ks} , \tau_{\rm p}=5\ {\rm d} , \dot T_{\rm obt} =-6.4\times10^{-6}$ ($\mathcal{H}_1$) and $ p= 248 \ M_\bullet, e= 0.01, T_{\rm obt}=17.4 \ {\rm ks} , \tau_{\rm p}=10\ {\rm d} , \dot T_{\rm obt,max} =-1.4\times10^{-5}$ ($\mathcal{H}_2$). 
In $\mathcal{H}_2$, the non-uniform orbital decay rate $\dot T_{\rm obt}$ is proportional to $z_{\rm disk}(t)\approx p \sin\iota_{\rm sd}(t)$ and the best-fit trajectory favors a low
$\dot T_{\rm obt}$ around XMM3. In $\mathcal{H}_1$, the apparent non-uniform orbital decay rate is the result of the disk alignment process.
The best-fit trajectory favors a fast disk alignment, where the apparent orbital period is longer than the true period during XMM1 due to the disk precession,
and converges to the true period after the disk alignment (XMM2-4).

The posterior corner plots of all model parameters are shown in Fig.~\ref{fig:eRO2_v_corner}, Fig.~\ref{fig:eRO2_m_corner} and Fig.~\ref{fig:eRO2_h2_corner}, respectively, where the orbital parameters are constrained as
\be 
\begin{aligned}
    p &= 502^{+396}_{-313}  \ M_\bullet\ , \\ 
    e &= 0.03^{+0.06}_{-0.03}  \ ,\quad (\mathcal{H}_0) \\
    T_{\rm obt} &= 17.1^{+0.2}_{-0.2} \ {\rm ks}\ ,
\end{aligned}
\ee 
and 
\be 
\begin{aligned}
    p &= 506^{+407}_{-286} \ M_\bullet\ , \\ 
    e &= 0.01^{+0.02}_{-0.01} \ , \\
    T_{\rm obt} &= 17.0^{+0.3}_{-0.2} \ {\rm ks}\ , \quad (\mathcal{H}_1)\\ 
    \tau_{\rm p} &= 7^{+24}_{-2} \ {\rm days} \ , \\
    \dot T_{\rm obt} &=-0.7^{+0.3}_{-0.4}\times10^{-5}\ ,
\end{aligned}
\ee 
and 
\be 
\begin{aligned}
    p &= 428^{+446}_{-221} \ M_\bullet\ , \\ 
    e &= 0.01^{+0.02}_{-0.01} \ , \\
    T_{\rm obt} &= 17.4^{+0.1}_{-0.1} \ {\rm ks}\ , \quad (\mathcal{H}_2)\\ 
    \tau_{\rm p} &= 10^{+28}_{-6} \ {\rm days} \ , \\
    \dot T_{\rm obt,max} &=-1.6^{+0.5}_{-1.0}\times10^{-5}\ ,
\end{aligned}
\ee 
at $95\%$ confidence level. 
The log Bayes factors between the three hypotheses are found to be
\be
\begin{aligned}
    \log\mathcal{B}^1_0 &=15.0\pm0.2 \ , \\ 
    \log\mathcal{B}^2_0 &=14.4\pm0.2 \ .
\end{aligned}
\ee 
Comparing with the vanilla hypothesis $\mathcal{H}_0$,  $\mathcal{H}_1$
and $\mathcal{H}_2$ are nearly equally favored, where the apparent orbital period decay 
is the result of disk alignment process in   $\mathcal{H}_1$ and is the true
orbital decay in $\mathcal{H}_2$. In spite of the difference in the two hypotheses, 
the orbital parameter constraints are consistent.

With the current data, we find the constraints on the SMBH mass as 
\be \label{eq:ero2_mass}
\begin{aligned}
    \mathrm{log}_{10}(M_{\bullet}/M_{\odot}) &= 4.7^{+0.5}_{-0.4}\ , \ (\mathcal{H}_1) \\
\mathrm{log}_{10}(M_{\bullet}/M_{\odot}) &= 4.8^{+0.5}_{-0.5}\ , \ (\mathcal{H}_2)
\end{aligned}
\ee 
at
2-$\sigma$ confidence level. These constraints are consistent with 
the value $\mathrm{log}_{10}(M_{\bullet}/M_{\odot}) =4.96\pm 1.1$  (2-$\sigma$) inferred from the $M_\bullet-\sigma_\star$ relation~\citep{Wevers2022}. Similar to the GSN 069 case, the SMBH mass obtained here is not entirely independent of the external measurement.
It is interesting to note that the SMBH of eRO-QPE2 is on the edge of intermediate mass BHs.
With longer monitoring of eRO-QPE2, we expect to distinguish the two  hypotheses and further pin down the SMBH mass.

Similar to in GSN 069, we can infer the orbital energy loss 
  per collision
\be 
 E_{\rm col} = \frac{E_{\rm obt}}{3} \dot T_{\rm obt} = 4.2 ^{+7.3}_{-2.0}\times 10^{45}\ \left(\frac{m}{M_\odot}\right)\   \mathrm{ergs} \ ,
\ee 
where we have used $\mathcal{H}_1$ as an example.
In combination with QPE energy $E_{\rm QPE}\approx 10^{45}$ ergs~\citep{Arcodia:2024taw},
we find the QPE radiation efficiency in eRO-QPE2
\be 
\eta_{\rm QPE} = \frac{E_{\rm QPE}}{E_{\rm col}} \approx 24^{+21}_{-15} \% \times \left(\frac{m}{M_{\odot}}\right)^{-1}\ ,
\ee 
at $95\%$ confidence level.
In the similar way, we can infer the disk surface density and the value of the $\alpha$ viscosity parameter.

Another feature in eRO-QPE2 which has been noticed recently \citep{Arcodia:2024taw} is the apparent orbital eccentricity evolution. In our EMRI trajectory model, the orbital eccentricity is a constant, therefore, the apparent orbital eccentricity change is a result of the relative orientation between the orbit and the disk. More detailed discussion can be found in \cite{Arcodia:2024taw}.

The orbital eccentricity evolution can be naturally incorporated into our EMRI trajectory model with a more physical modeling 
of the EMRI-disk collisions. For the purpose of this work, this is not necessary, because the small eccentricity change ($\delta e\sim e \times \delta T_{\rm obt}/T_{\rm obt}$) during 
the observation period is beyond the detection limit of the data available.

\section{Summary and discussions}\label{sec:conclusions}

\subsection{Summary}
In this work, we have extended the previous orbital analyses of QPE timing by including the EMRI orbital period decay due to collisions with the accretion disk
and the disk precession/alignment if the disk is initially misaligned. We have applied this generalized orbital analysis to GSN 069 and eRO-QPE2,
the two most stable  QPE sources so far.

Regarding GSN 069, we find clear Bayesian evidence for the non-zero orbital period decay rate $\dot T_{\rm obt}$ [Eq.~(\ref{eq:GSN_cons_H1})], 
along with tight constraints on the SMBH mass $M_\bullet$, the orbital size $p$ and the orbital eccentricity $e$.
From these constraints, we can directly measure the EMRI orbital energy loss as crossing the accretion disk $E_{\rm col}$, and the QPE radiation efficiency 
$\eta_{\rm QPE}$ in combination with the QPE light curves. We find $\eta_{\rm QPE}\approx 10\% (M_\odot/m)$ in the case of GSN 069. This radiation efficiency 
looks reasonable for a normal star with $m\approx 1 M_\odot$ as the secondary object
but seems to too low for a sBH with $m \gtrsim 30 M_\odot$. A firm conclusion on the nature of the SMO based on the measured QPE radiation efficiency 
requires detailed analysis of the QPE radiation processes.

Regarding eRO-QPE2, the authors of \citealt{Arcodia:2024taw} already noticed its non-uniform orbital decay and interpreted this behavior as a result of the disk precession [Eq.~(\ref{eq:vary_Tdot})].
In this work, we consider another possibility: the apparent non-uniform orbital decay is a result of the disk alignment process.
In the framework of Bayesian analysis on the QPE timing, we find the two hypotheses are nearly equally favored by the current data. 
The two hypotheses predict distinct future evolution of the EMRI orbital period: the former predicts a periodic change in $\dot T_{\rm obt}$,
while the latter predicts a constant  $\dot T_{\rm obt}$. The two different predictions should be able to be distinguished by the 2024 observations of XMM-Newton.
We will do a follow-up analysis when the 2024 data are released for public use.

With the generalized QPE timing model, we incorporate different observations spanning a long time into  a single orbital analysis and 
obtain tighter constraints on the EMRI orbital parameters. The tighter constraints of the orbital eccentricities further confirm
the conclusion on the EMRI formation history in papers I and II: they are consistent with the wet channel prediction~\citep{Sigl2007,Levin2007,Pan2021prd,Pan2021b,Pan2021,Pan2022,Derdzinski2023,Wang2023,Wang2023b}, 
but incompatible with either the dry channel~\citep{Hopman2005,Preto2010,Bar-Or2016,Babak2017,Amaro2018,Broggi2022} or the Hills channel~\citep{Miller2005,Raveh2021}.

\begin{figure}
\includegraphics[scale=0.32]{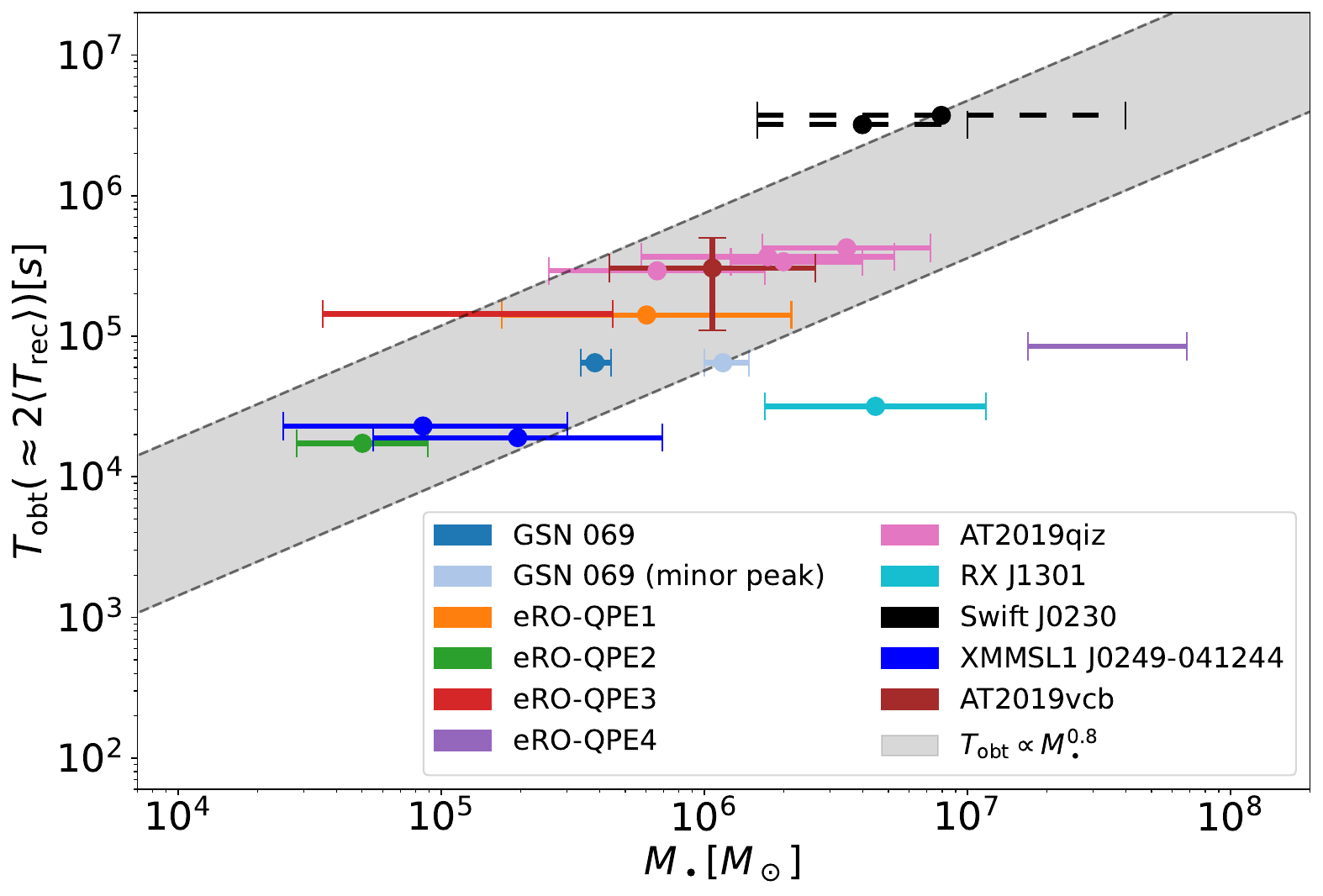}
\caption{\label{fig:T_vs_M} Population properties of QPE sources: Orbital period ($T_{\rm obt}\approx 2\langle T_{\rm rec} \rangle$) versus SMBH mass ($M_\bullet$). Error bars indicate 1-$\sigma$ uncertainties. We include four independent mass measurements for AT2019qiz (from TDE disk modeling and three $M_\bullet-\sigma_\star$ relations~\citep{nicholl2020,nicholl2024}), two for Swift J0230 (derived from different host galaxy scaling relations~\citep{Guolo2024}), two for XMMSL1 J0249-041244 (from two $M_\bullet-\sigma_\star$ relations~\citep{Wevers2019,Wevers2022}). Dashed error bars denote Swift J0230, reflecting its uncertain identification as a QPE source.}
\end{figure}

Tighter constraints of the orbital parameters also open the possibility of accurately quantifying the population statistics of QPE sources.
In Fig.~\ref{fig:T_vs_M}, we show the orbital periods $T_{\rm obt}$ and the SMBH masses $M_\bullet$ of 9 QPE sources and 
Swift J0230 which has not been confidently identified as a QPE source since its excpetional long recurrence times and 
opposite spectral evolution during the flares \citep{Evans2023,Guolo2024}. The SMBH masses of GSN 069 and eRO-QPE2 are inferred 
from the QPE timing as shown in this paper [Eqs.~(\ref{eq:mass_gsn},\ref{eq:ero2_mass})], and others are from host galaxy scaling relations~\citep{Wevers2019,Wevers2022, Guolo2024, Arcodia2024, nicholl2020, Wevers2019, Yao_2023, Wevers:2024hzp} or TDE disk modeling~\citep{nicholl2024}. For sources lacking a measured $T_{\rm obt}$, we adopt twice the average recurrence time
$\langle T_{\rm rec}\rangle$ as a reliable proxy \citep{Chakraborty2021,jiang2025embersactivegalacticnuclei}. 
The QPE recurrence time $\langle{T}_{\rm rec}\rangle$ of AT~2019vcb is most uncertain due to lacking consecutive flares detected
(see \cite{jiang2025embersactivegalacticnuclei} for details of the estimate).
Similar diagram could be also be found in \citealt{Guolo2024,nicholl2024}, where no clear evidence for the $T_{\rm obt}-M_\bullet$ correlation was found.
With better constraints of the SMBH masses of GSN 069 and eRO-QPE2, we find a  likely correlation $T_{\rm obt}\propto M_\bullet^{n}$ 
with $n\approx 0.8$,
excluding RX J1301 and eRO-QPE4. These two QPE sources are exceptional in their high orbital eccentricities (see \citealt{Franchini2023,Zhou2024b,Giustini:2024dyy} for detailed analyses of RX J1301 and \citealt{Arcodia2024} for the light curves of eRO-QPE4 which shows a large difference between the long recurrence times and the short ones).
Therefore, an interesting implication of this diagram is that there are two populations of QPEs sourced by low-eccentricity and high-eccentricity EMRIs, respectively.
The low-eccentricity population follows a scaling $T_{\rm obt}\propto M_\bullet^{0.8}$, while it is unclear whether a similar correlation exists in the high-eccentricity population due to the limited number of sources. If the dichotomy is confirmed with more QPE sources to be discovered in the future, 
it will be a smoking-gun signature of multiple EMRI formation channels and the correlation(s) further shed light on the 
physical processes functioning in different channels.

\subsection{Comparison with previous work}

As we were finishing this paper (arXiv:2411.18046),    arXiv:2411.13460 appeared \citep{Miniutti:2024rlj}, where the authors did a 
detailed Observed minus Calculated  (O-C) analysis of the GSN 069 QPE timing, 
and found evidence for orbital period decay and for super-orbital modulation on tens of days. 

With  observations XMM 1-3 and Chandra (Dec. 2018- May 2019), they found three possible solutions to the 
orbital period decay rate: $\dot P_{\rm obt}^{(2019)} = 0, -(3, 4)\times 10^{-5}$ or $-(6, 7)\times 10^{-5}$, where the third solution is nicely 
consistent with our result $\dot T_{\rm obt}^{(2019)}=-(6.5\pm 0.3)\times 10^{-5}$ [Eq.~(\ref{eq:GSN_cons_H1})]. After May 2019, the QPEs became irregular,
then disappeared for a number of years along with the rebrightening of the background emission, which is believed to be the signature of the second partial TDE~\citep{Miniutti2023}. The QPE disappearance is likely the result of disk destruction by the second partial TDE, and the 
 reappearance of regular QPEs a few years later is a signature of a new disk formation. 
New QPEs were indeed found in the 2023 observations as recently reported by \citealt{Miniutti:2024rlj}. Combining the old 2019 data and the new 2023 data,
they reported an average decay rate during 2019-2023, $\dot P_{\rm obt}^{(2019-2023)}= -(3.7\pm 1.3)\times 10^{-5}$.
This does NOT mean the second solution to the orbital decay rate during Dec. 2018 - May 2019, $\dot P_{\rm obt}^{(2019)}=-(3, 4)\times 10^{-5}$ is favored.
The disk structure had undergone a substantial disruption during 2019-2023, and the disk may even have  been destructed completely during the QPE disappearance. Therefore, a slower orbital decay during this period (2019-2023) than in the regular phase (Dec. 2018-May 2019) is expected, i.e., $|\dot P_{\rm obt}^{(2019-2023)}| <|\dot P_{\rm obt}^{(2019)}|$.

\begin{figure*}
\includegraphics[scale=0.4]{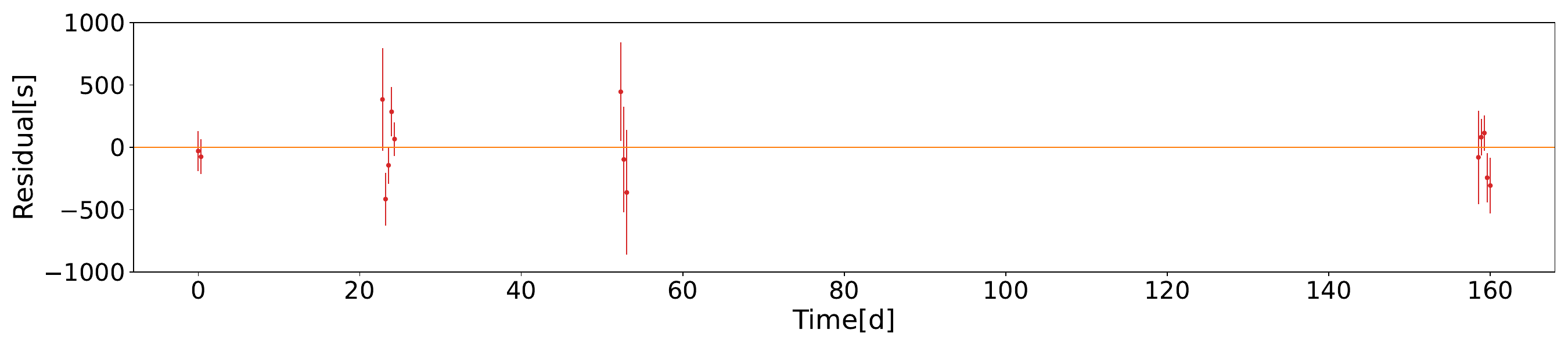}
\caption{\label{fig:Res_gsn} Residuals of the best-fit QPE timing model with $\mathcal{H}_1$ for GSN 069, i.e. $t_{\rm obs}^{(k)}-t_0^{(k)}$,
where error bars represent 1-$\sigma$ uncertainties.}
\end{figure*}

\citealt{Miniutti:2024rlj} also found two periods in the QPE timing data, one is the orbital period  and 
the other super-orbital period ($P_{\rm mod}\sim 19 $ d) was interpreted as evidence for  disk precession or a sub-milliparsec SMBH binary.
But we found no evidence for a third period other than the orbital period $T_{\rm obt}\approx 64$ ks and the apsidal precession period $T_{\rm prec}\approx 102 T_{\rm obt}$
(see the red line in Fig.~\ref{fig:lo_GSN} for a reasonable fit to the data without invoking disk precession and the featureless 
residuals in Fig.~\ref{fig:Res_gsn}).
The discrepancy may come from different models and different analysis methods used. We fit the QPE timing data with forced EMRI trajectories in the Kerr spacetime, 
and all the data interpretation is done in the Bayesian analysis framework.
\citealt{Miniutti:2024rlj} adopted a simple phenomenological model instead and interpreted the data in a pictorial approach.
Note that their best-fit sinusoidal modulation period $P_{\rm mod}\sim 19 $ d is coincident with $1/4$ times the median value of the  EMRI apsidal precession period in our orbital analysis,
 $T_{\rm prec}/4 = 102 T_{\rm obt}/4 =  19.1$ d [Eq.~(\ref{eq:GSN_cons_H1})]. 
This coincidence indicates that  a similar  super-orbital period in the QPE timing was found in these two works, 
but was interpreted  in different ways: 
intrinsic apsidal precession + no external source (this work)
v.s. an external source + no intrinsic apsidal precession \citep{Miniutti:2024rlj}. 
\cite{Miniutti:2024rlj} disfavored the apsidal precession 
because apsidal precession causes the even and odd QPE timings to modulate in anti-phase, 
whereas the observed timings with their O-C analyses seem to modulate nearly in-phase. 
In our work, a direct comparison with EMRI trajectory predictions with the QPE timing data in the Bayesian inference framework indeed reveals a best-fit model that nicely and naturally explains the current data (see Figs.~\ref{fig:lo_GSN} and \ref{fig:Res_gsn}).
It is out of the scope of this work to figure out the origin of the discrepancy.

\begin{acknowledgments}
We thank Ning Jiang and Shifeng Huang for valuable discussions. The computations in this paper were run on the Siyuan-1 cluster supported by the Center for High Performance Computing at Shanghai Jiao Tong University.
\end{acknowledgments}

\software{bilby \citep{Ashton2019},  
          nessai \citep{nessai}, 
          pymultinest \citep{pymultinest}
          }

\appendix

\begin{figure*}
\includegraphics[scale=0.31]{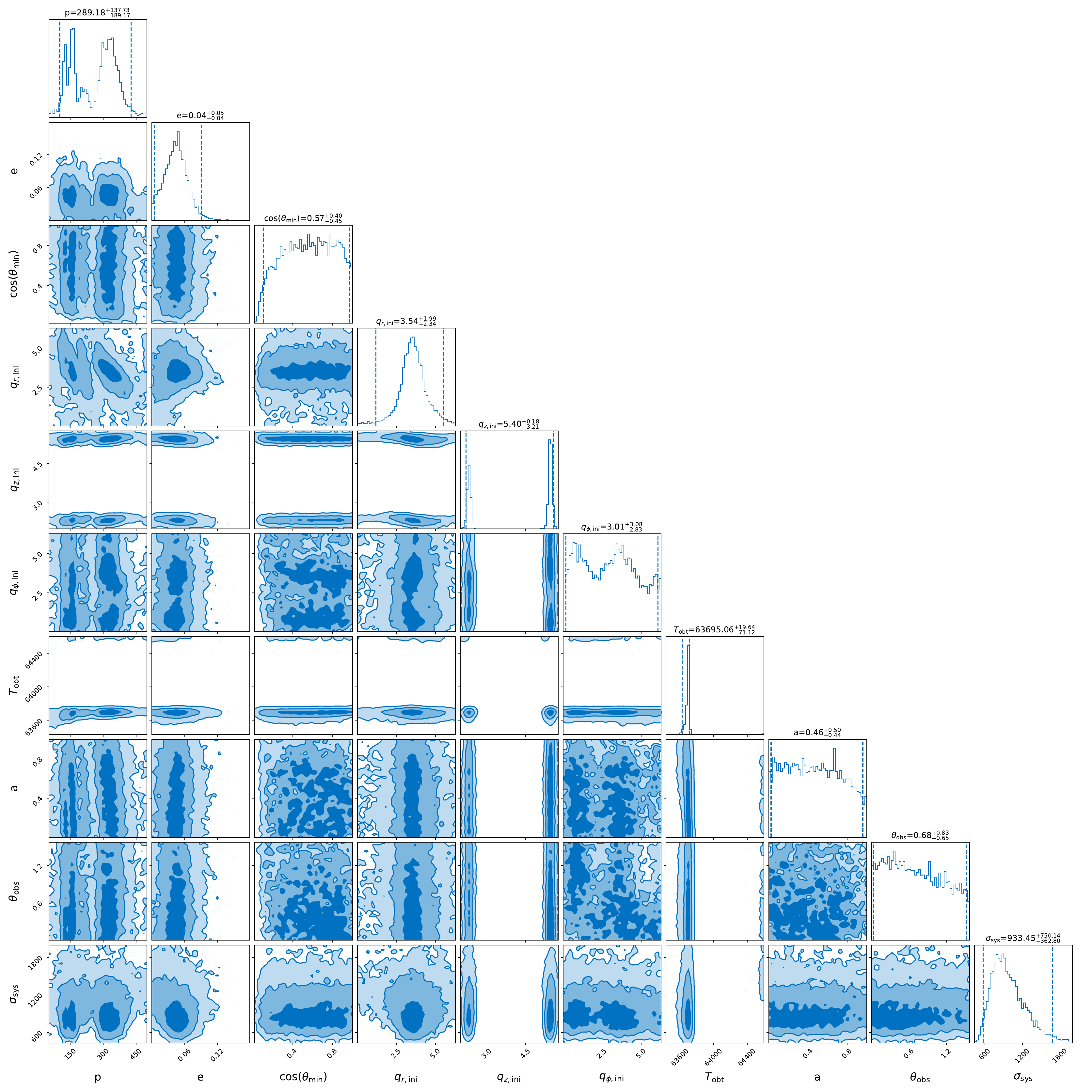}
\caption{\label{fig:GSN069_H0} The posterior corner plot of model parameters for GSN 069 
 with the vanilla hypothesis ($\mathcal{H}_{0}$) : $p [M_\bullet], e, 
\cos\theta_{\rm min}, q_{r,\mathrm{ini}}, q_{z,\mathrm{ini}}, q_{\phi,\mathrm{ini}},
T_{\rm obt} [{\rm sec}], a, \theta_{\rm obs}, \sigma_{\rm sys} [{\rm sec}]$, where each pair of vertical lines denotes the 2-$\sigma$ confidence level.}
\end{figure*}

\begin{figure*}
\includegraphics[scale=0.31]{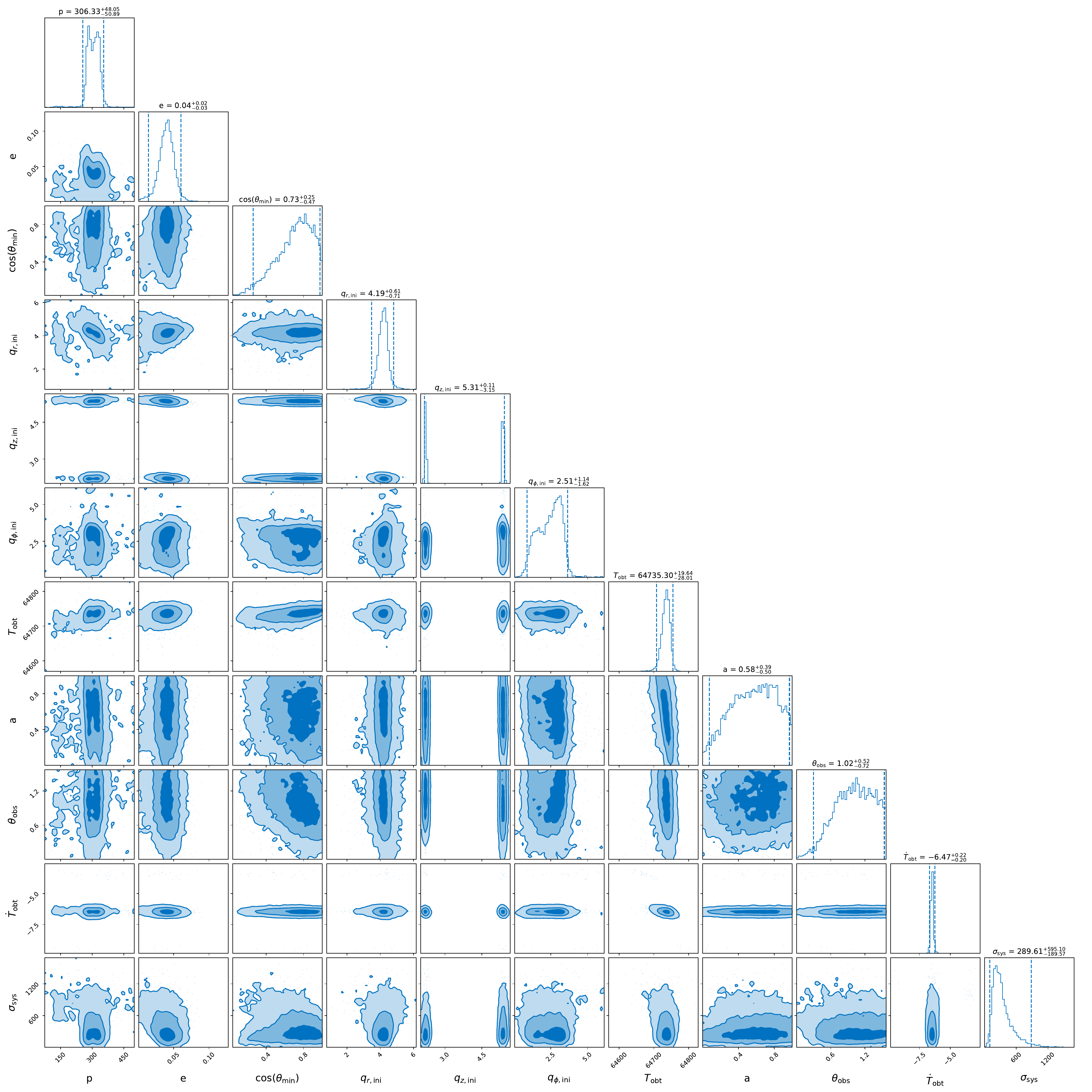}
\caption{\label{fig:GSN069_H1} The posterior corner plot of model parameters for GSN 069 
 with a forced EMRI + an equatorial disk hypothesis ($\mathcal{H}_{1}=\mathcal{H}_{\rm e1}+\mathcal{H}_{\rm d0}$) : $p [M_\bullet], e, 
\cos\theta_{\rm min}, q_{r,\mathrm{ini}}, q_{z,\mathrm{ini}}, q_{\phi,\mathrm{ini}},
T_{\rm obt} [{\rm sec}], a, \theta_{\rm obs}, \dot{T}_{\rm obt} [\times10^{-5}], \sigma_{\rm sys} [{\rm sec}]$, where each pair of vertical lines denotes the 2-$\sigma$ confidence level.}
\end{figure*}

\begin{figure*}
\includegraphics[scale=0.31]{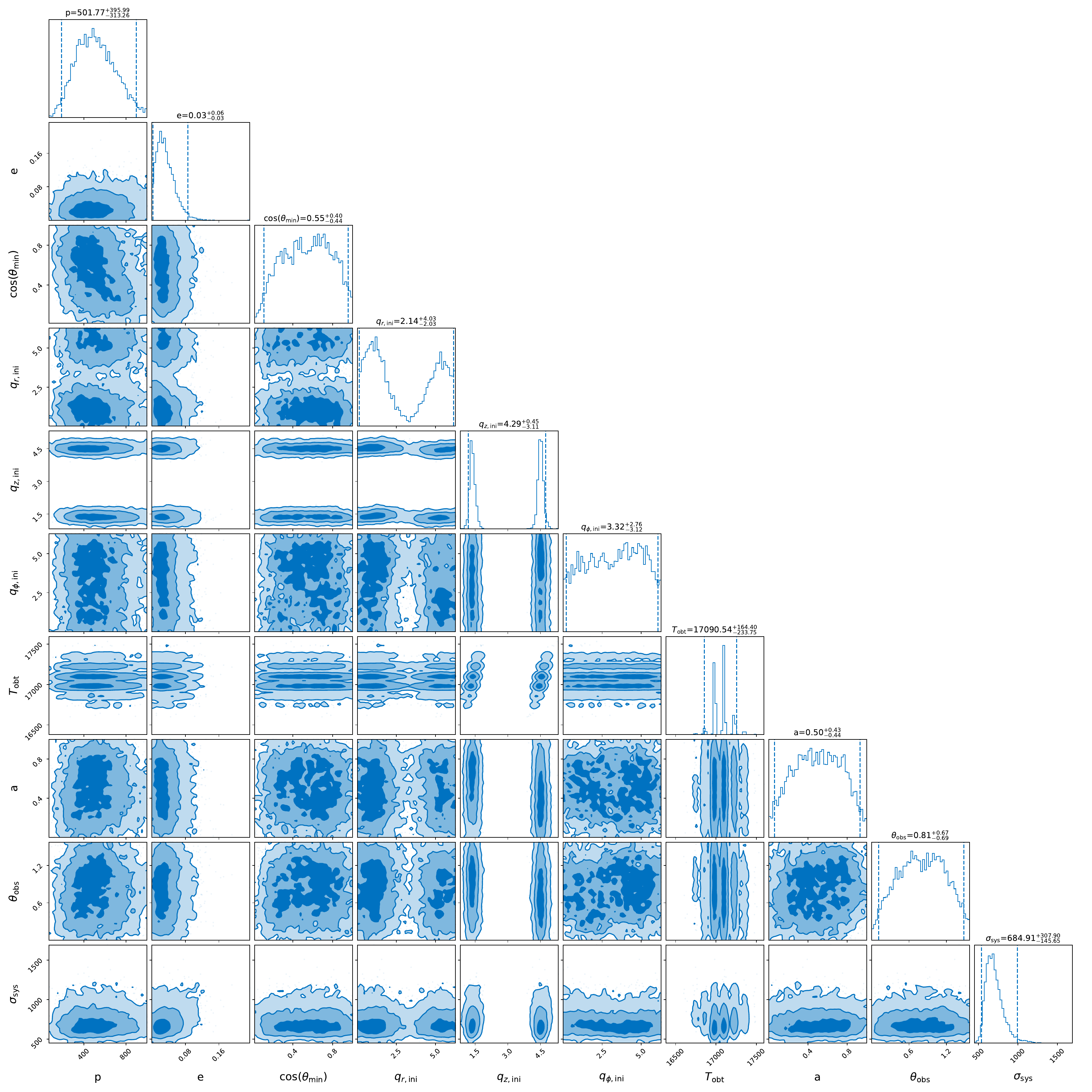}
\caption{\label{fig:eRO2_v_corner} Same to Fig.~\ref{fig:GSN069_H0} except for eRO-QPE2.}
\end{figure*}

\begin{figure*}
\includegraphics[scale=0.25]{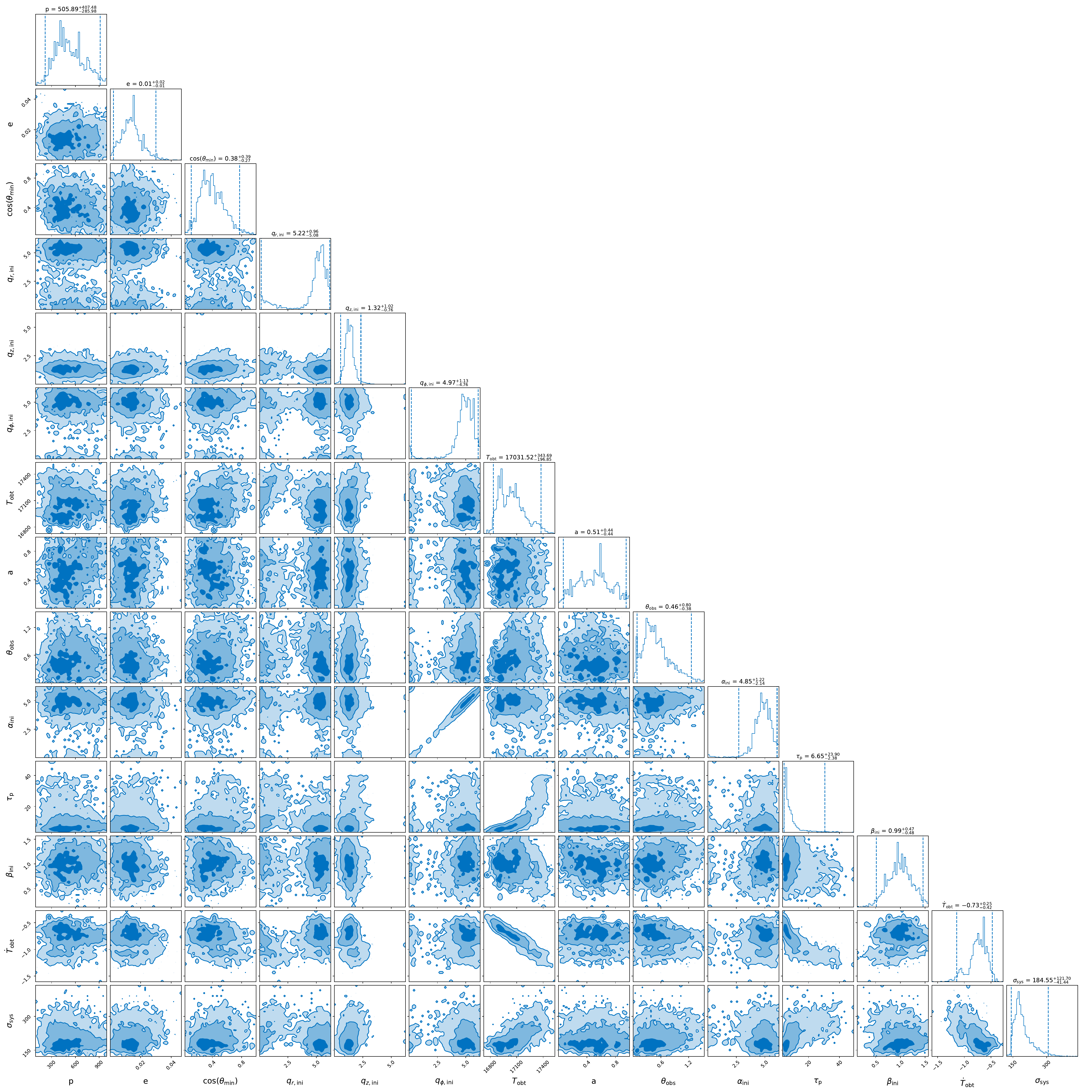}
\caption{\label{fig:eRO2_m_corner}  The posterior corner plot of model parameters for eRO-QPE2
 with the hypothesis of an initially misaligned disk ($\mathcal{H}_1=\mathcal{H}_{\rm e1}+\mathcal{H}_{\rm df}$): $p [M_\bullet], e, 
\cos\theta_{\rm min}, q_{r,\mathrm{ini}}, q_{z,\mathrm{ini}}, q_{\phi,\mathrm{ini}},
T_{\rm obt} [{\rm sec}], a, \theta_{\rm obs}, \alpha_{\rm ini}, \tau_{\rm p} [{\rm days}], \beta_{\rm ini}, \dot{T}_{\rm obt} [\times10^{-5}], \sigma_{\rm sys} [{\rm sec}]$, where each pair of vertical lines denotes the 2-$\sigma$ confidence level.}
\end{figure*}

\begin{figure*}
\includegraphics[scale=0.25]{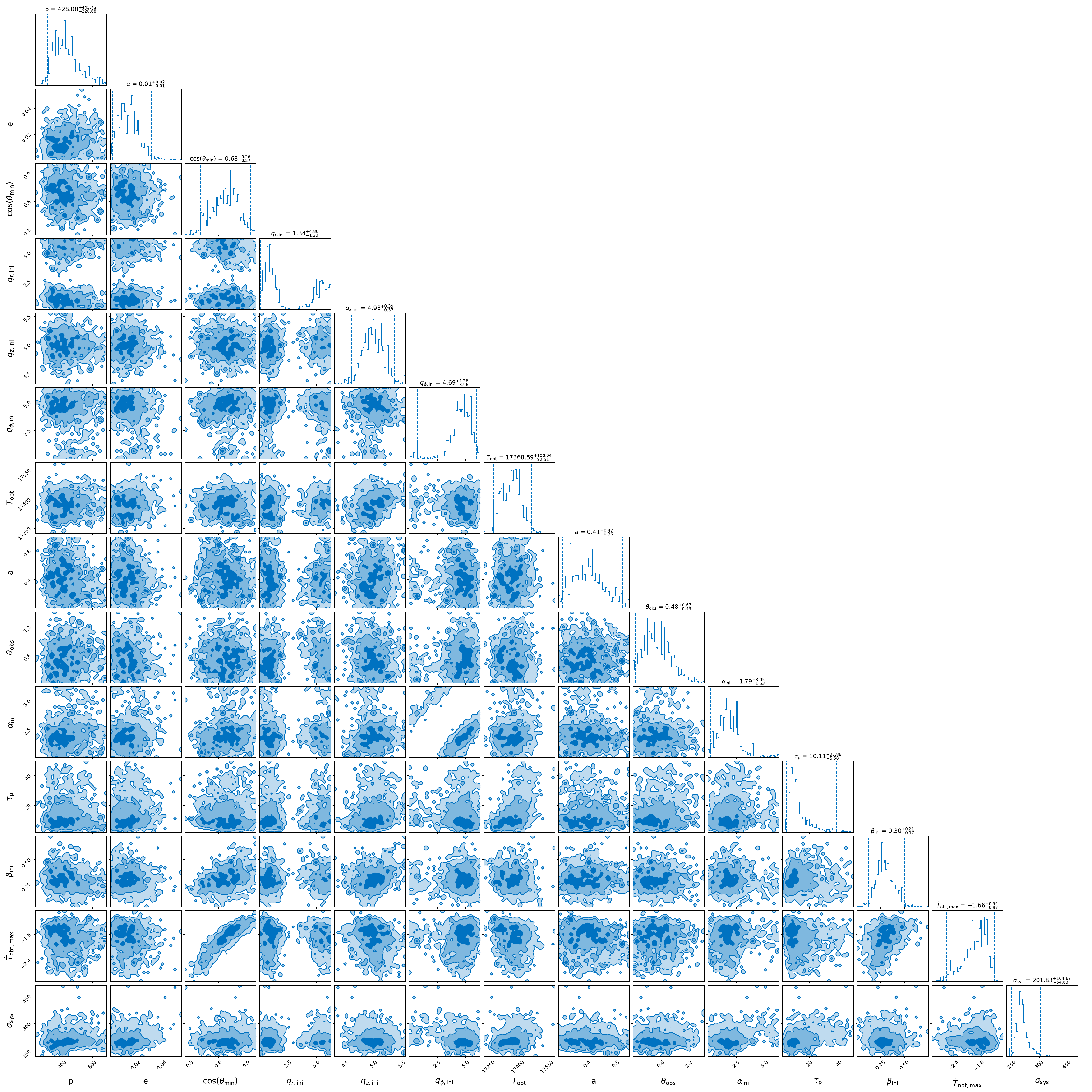}
\caption{\label{fig:eRO2_h2_corner}  The posterior corner plot of model parameters for eRO-QPE2
 with the hypothesis of an nonuniform orbital decay and a precessing disk ($\mathcal{H}_2=\mathcal{H}_{\rm e1}+\mathcal{H}_{\rm ds}$): $p [M_\bullet], e, 
\cos\theta_{\rm min}, q_{r,\mathrm{ini}}, q_{z,\mathrm{ini}}, q_{\phi,\mathrm{ini}},
T_{\rm obt} [{\rm sec}], a, \theta_{\rm obs}, \alpha_{\rm ini}, \tau_{\rm p} [{\rm days}], \beta_{\rm ini}, \dot{T}_{\rm obt, max} [\times10^{-5}], \sigma_{\rm sys} [{\rm sec}]$, where each pair of vertical lines denotes the 2-$\sigma$ confidence level.}
\end{figure*}

\bibliography{ms}{}
\bibliographystyle{aasjournal}

\end{document}